\documentclass{siamltex}
\usepackage{euscript,amsmath,amssymb,amsfonts,graphicx,bm}
  \usepackage{paralist}
  \usepackage{epstopdf}
  \usepackage{graphics} 

\newcommand{\n}{{\mathbf n}}
\newcommand{\m}{{\mathbf m}}
\newcommand{\e}{{\rm e}}
\renewcommand{\d}{{\rm d}}

\newcommand{\vphi}{{\bm \phi}_{\epsilon}}

\newcommand{\p}{{\bf p}}

\renewcommand{\u}{{\mathbf u}}

\newcommand{\Z}{{\mathbf Z}}

\newcommand{\x}{{\mathbf x}}

\title{Metastability in a stochastic neural network modeled as a velocity jump Markov process\thanks{PCB was supported
by the National Science Foundation (DMS-1120327) and JMN by the NSF funded Mathematical Biosciences Institute.} }

\author{Paul C. Bressloff\thanks{Department of Mathematics, University of Utah, Salt Lake City, UT 84112 USA  ({\tt bressloff@math.utah.edu})} \and Jay M. Newby\thanks{Mathematical Biosciences Institute, Ohio State University ({\tt newby@math.utah.edu})}} 

\date{today}

\begin{document}

\maketitle
\newcommand{\slugmaster}{%
\slugger{MMedia}{xxxx}{xx}{x}}

\begin{abstract}

One of the major challenges in neuroscience is to determine how noise that is present at the molecular and cellular levels affects dynamics and information processing at the macroscopic level of synaptically coupled neuronal populations. Often noise is incorprated into deterministic network models using extrinsic noise sources. An alternative approach is to assume that noise arises intrinsically as a collective population effect, which has led to a master equation formulation of stochastic neural networks. In this paper we extend the master equation formulation by introducing a stochastic model of neural population dynamics in the form of a velocity jump Markov process. The latter has the advantage of keeping track of synaptic processing as well as spiking activity, and reduces to the neural master equation in a particular limit. The population synaptic variables evolve according to piecewise deterministic dynamics, which depends on population spiking activity. The latter is characterised by a set of discrete stochastic variables evolving according to a jump Markov process, with transition rates that depend on the synaptic variables. We consider the particular problem of rare transitions between metastable states of a network operating in a bistable regime in the deterministic limit. Assuming that the synaptic dynamics is much slower than the transitions between discrete spiking states, we use a WKB approximation and singular perturbation theory to determine the mean first passage time to cross the separatrix between the two metastable states. Such an analysis can also be applied to other velocity jump Markov processes, including stochastic voltage-gated ion channels and stochastic gene networks.

\end{abstract}



\pagestyle{myheadings}
\thispagestyle{plain}
\markboth{P. C. BRESSLOFF AND J. M. NEWBY}{NEURAL VELOCITY JUMP MARKOV PROCESS}

\normalsize

\section{Introduction}

Noise has recently emerged as a key component of many biological systems including the brain. Stochasticity arises at multiple levels of brain function, ranging from molecular processes such as gene expression and the opening of ion channel proteins to complex networks of noisy spiking neurons that generate behaviour \cite{Faisal08}. For example, the spike trains of individual cortical neurons {\em in vivo} tend to be very noisy, having interspike interval (ISI) distributions that are close to Poisson \cite{Softky92}. At the network level, noise appears to be present during perceptual decision making \cite{wang08} and bistable perception, the latter being exemplified by perceptual switching during binocular rivalry \cite{morenobote07,shpiro09,Webber12}. Noise also contributes to the generation of spontaneous activity during resting states \cite{Destexhe06,Deco11}. At the level of large-scale neural systems, as
measured with functional MRI (fMRI) imaging, this ongoing spontaneous activity reflects the organization of a series of highly coherent functional networks that may play an important role in cognition. One of the major challenges in neuroscience is to develop our understanding of how noise that is present at the molecular and cellular levels affects dynamics and information processing at the macroscopic level of synaptically coupled neuronal populations. Mathematical and computational modeling are playing an increasing role in developing such an understanding \cite{Laing09}.

Following studies of biochemical and gene networks \cite{Swain06,Faisal08}, it is useful to make the distinction between intrinsic and extrinsic noise sources. Extrinsic noise refers to external sources of randomness associated with environmental factors, and is often modeled as a continuous Markov process based on Langevin equations. On the other hand, intrinsic noise typically refers to random fluctuations arising from the discrete and probabilistic nature of chemical reactions at the molecular level, which are particularly significant when the number of reacting molecules $N$ is small. Under such circumstances, the traditional approach to modeling chemical reactions based on the law of mass action is inappropriate. Instead, a master equation formulation is necessary in order to describe the underlying jump Markov process. In the case of single cortical neurons, the main source of extrinsic noise arises from synaptic inputs. That is, cortical neurons are being constantly bombarded by thousands of synaptic currents, many of which are not correlated with a meaningful input and can thus be treated as background synaptic noise. The main source of intrinsic fluctuations at the single cell level is channel noise, which arises from the variability in the opening and closing of a finite number of ion channels. The resulting conductance--based model of a neuron can be formulated as a stochastic hybrid system, in which a piecewise smooth deterministic dynamics describing the time evolution of the membrane potential is coupled to a jump Markov process describing channel dynamics \cite{Pakdaman10,Buckwar10,Keener11}.

It is not straightforward to determine how noise at the single cell level translates into noise at the population or network level. A number of methods involve carrying out some form of dimension reduction of a network of synaptically-coupled spiking neurons. These include population density methods \cite{Nykamp00,Omurtag00,Ly07}, mean field theories \cite{Abbott93,Gerstner93,Brunel99,Brunel00,Touboul12}, and Boltzmann--like kinetic theories \cite{Cai04,Rangan08,Buice11}. However, such methods tend to consider either fully-connected or sparsely connected networks and simplified models of spiking neurons such as the integrate-and-fire (IF) model. Nevertheless, one interesting result that emerges from the mean-field analysis of IF networks is that, under certain conditions, even though individual neurons exhibit Poisson--like statistics, the neurons fire asynchronously so that the total population activity evolves according to a mean--field rate equation with a characteristic activation or gain function \cite{Abbott93,Gerstner93,Brunel99,Brunel00}. Formally speaking, the asynchronous state only exists in the thermodynamic limit $N\rightarrow \infty$, where $N$ determines the size of the population. This then suggests a possible source of intrinsic noise at the network level arises from fluctuations about the asynchronous state due to finite size effects \cite{Meyer02,Mattia02,Soula07,Boustani09}; this is distinct from intrinsic noise at the single cell level due to channel fluctuations and it is assumed that the latter is negligible at the population level. The presence of finite-size effects has motivated developing a closer analogy between intrinsic noise in biochemical and neural networks \cite{Bressloff09,Bressloff10}, based on a rescaled version of the neural master equation introduced by Buice {\em et. al.} \cite{Buice07,Buice09}, see also \cite{Ohira97}. 

In the Buice {\em et al} master equation \cite{Buice07,Buice09}, neurons are partitioned into a set of $M$ local homogeneous populations. The state of the $\alpha$th population at time $t$ is specified by the number $N_{\alpha}(t)$ of active (spiking) neurons in an infinite background sea of inactive neurons. (This is reasonable if the networks are in low activity states). Transitions between the states are given by a one-step jump Markov process, with the transition rates chosen so that standard Wilson-Cowan or activity-based equations are obtained in the mean-field limit, where statistical correlations can be ignored. One of the features of the Buice {\em et. al.} master equation is that there does not exist a natural small parameter, so that it is not possible to carry out a diffusion-like approximation using, for example, a system-size expansion. Indeed, the network tends to operate in a regime close to Poisson-like statistics. Neverthless, it is possible to solve the moment hierarchy problem using either path-integral methods or factorial moments \cite{Buice07,Buice09}. In contrast, the Bressloff master equation \cite{Bressloff09,Bressloff10} assumes that there is a finite number $N$ of neurons in each local population and characterizes the state of each population in terms of the fraction of neurons $N_{\alpha}(t)/N$ that have spiked in an interval of width $\Delta t$. The transition rates are rescaled so that in the thermodynamic limit $N\rightarrow \infty$, one recovers the Wilson-Cowan mean-field equations. For large, but finite $N$, the network operates in a  Gaussian-like regime that can be described in terms of an effective neural Langevin equation \cite{Bressloff09,Bressloff10}. One of the advantages of this version of the master equation from a mathematical perspective, is that a variety of well-established methods from the analysis of chemical master equations can be generalized to the neural case. For example, a rigorous analysis of the Langevin approximation can be carried out \cite{Buckwar12} by extending the work of Kurtz \cite {Kurtz76} on chemical master equations. Moreover WKB methods can be used to analyze rare transitions between metastable states, for which the Langevin approximation breaks down \cite{Bressloff10}. For a discussion of the differences between the two master equations from the perspective of corrections to mean field theory, see \cite{Touboul11b}. 

In this paper we go beyond the neural master equations by formulating the network population dynamics in terms of a stochastic hybrid system described by a ``velocity'' jump Markov process. This generalization is motivated by a major limitation of the neural master equations. That is, they neglect synaptic dynamics completely, only keeping track of changes in spiking activity. This implies, for example, that the relaxation time $\tau$ for synaptic dynamics is much smaller than the fundamental time step $\Delta t$ for jumps in the number of active neurons. Our model associates with each population two stochastic variables $U_{\alpha}(t)$ and $N_{\alpha}(t)$. The synaptic variables $U_{\alpha}(t)$ evolve according to piecewise--deterministic dynamics describing, at the population level, synapses driven by spiking activity. These equations are only valid between jumps in spiking activity, which are described by a jump Markov process whose transition rates depend on the synaptic variables. Formally speaking, the resulting stochastic dynamics can be modeled in terms of a differential Chapman-Kolmogorov (CK) equation:
\begin{equation}
\label{0}
\frac{\partial p}{\partial t}=-\frac{1}{\tau}\sum_{{\alpha}=1}^M\frac{\partial}{\partial u_{\alpha}}(v_{\alpha}(\u,\n)p(\u,\n,t))+\frac{1}{\tau_a}\sum_{\m}A(\n,\m;\u)p(\u,\m,t).
\end{equation}
Here $\n=(n_1,\ldots,n_M)$, $\u=(u_1,\ldots,u_M)$, and $p(\u,\n,t)$ is the state probability density at time $t$. The drift ``velocities'' $v_{\alpha}(\u,\n)$ for fixed $\n$ represent the piecewise-deterministic synaptic dynamics according to
\begin{equation}
\tau \frac{du_{\alpha}}{dt}=v_{\alpha}(\u,\n),\quad \alpha=1,\ldots,M,
\end{equation}
and $A$ represents the $\u$--dependent transition matrix for the jump Markov process with $\sum_{\n}A(\n,\m;\u)=0$ for all $\m$. Note that the transition rates are scaled by a second time constant $\tau_a$ that characterizes the relaxation rate of population activity. In the limit $\tau\rightarrow 0$ for fixed $\tau_a$, equation (\ref{0}) reduces to the neural master equation \cite{Buice07,Buice09,Bressloff09} with $\u=\u(\n)$ such that $v_{\alpha}(\u(\n),\n)=0$. On the other hand, if $\tau_a\rightarrow 0$ for fixed $\tau$, then we obtain deterministic voltage or current-based mean-field equations
\begin{equation}
\tau \frac{du_{\alpha}}{dt}=v_{\alpha}(\u,\langle \n\rangle ),
\end{equation}
where $\langle \n\rangle =\sum_{\n}\rho(\u,\n)\n$ with $\rho(\u,\n)$ the steady-state density satisfying the equation
$\sum_{\m}A(\n,\m;\u)\rho(\u,\m)=0$.
It is straightforward to show using the Perron-Frobenius Theorem that the steady-state density exists and is unique. Note that the limit $\tau_a\rightarrow 0$ is analogous to the slow synapse approximation used by Ermentrout \cite{Ermentrout94} to reduce deterministic conductance-based neuron models to voltage-based rate models.
Now suppose that the network operates in the regime $0<\tau_a/\tau\equiv\epsilon \ll 1$, for which there are typically a large number of transitions between different firing states $\n$ while the synaptic currents $\u$ hardly change at all. This suggests that the system rapidly converges to the (quasi) steady state $\rho(\u,\n)$, which will then be perturbed as $\u$ slowly evolves. The resulting perturbations can be analyzed using a quasi-steady-state (QSS) diffusion or adiabatic approximation \cite{Papanicolaou,Gardiner09,Newby10a}, in which the CK equation (\ref{0}) is approximated by a Fokker-Planck equation. The latter captures the Gaussian-like fluctuations within the basin of attraction of a fixed point of the mean-field equations, and can be used to investigate effects such as the noise-induced amplification of subthreshold oscillations (quasicycles) along smililar lines to \cite{Bressloff10}. However, the diffusion approximation for small $\epsilon$ breaks down when considering rare event transitions between metastable states. (A similar problem arises in approximating chemical master equations by a Fokker--Planck equation in the large $N$ limit \cite{Hanggi84,Dykman94}).

In this paper we will show how asymptotic methods recently developed to study metastability in stochastic ion channels and gene networks  \cite{Keener11,Newby12,Newby12a} can be extended to analyze metastability in stochastic neural networks. All of these systems are modeled in terms of a stochastic hybrid system evolving according to a CK equation of the form (\ref{0}). For example, in the case of ion channels, $n_{\alpha}$ would represent the number of open channels of type $\alpha$, whereas $\u$ would be replaced by the membrane voltage $v$. The neural system is distinct, in that the numbers of discrete and continuous variables are equal. The structure of the paper is as follows. In \S 2, we present our stochastic network model and the associated neural CK equation, and carry out the QSS diffusion approximation for small $\epsilon$. We then analyze bistability in a one-population model (\S 3), and a two-population model consisting of a pair of excitatory and inhibitory networks (\S 4). In both cases, we carry out an eigenfunction expansion of the probability density and equate the principal eigenvalue with the inverse mean first pasage time from one metastable state to the other. The principal eigenvalue is expressed in terms of the inner product of a quasistationary density and an adjoint eigenfunction. The former is evaluated using a WKB approximation, whereas the latter is determined using singular perturbation theory, in order to match an absorbing boundary condition on the separatrix between the basins of attraction of the two metastable states. In the two-population case, calculating the effective potential of the quasistationary density requires identifying an appropriate Hamiltonian system, which turns out to be non-trivial, since the system does not satisfy detailed balance.

A number of general comments are in order before proceeding further.
\vspace{0.1cm}

\noindent (i) There does not currently exist a rigorous derivation of population rate-based models starting from detailed biophysical models of individual neurons. Therefore, the construction of the stochastic rate-based model in \S 2 is heuristic in nature, in order to motivate the neural rate equations used in this paper.
\medskip

\noindent (ii) We use formal asymptotic methods rather than rigorous stochastic analysis to determine the transition rates between metastable states in \S 3 and \S 4, and validate our approach by comparing with Monte-Carlo simulations. Such methods have been applied extensively to Fokker-Planck equations and master equations as reviewed in \cite{Schuss10}, and provide useful insights into the underlying dynamical processes. In this paper we extend these methods to a stochastic hybrid system. One could develop a more rigorous approach using large deviation theory \cite{Freidlin98,Touchette09}, for example, although as far as we are aware this has not been fully developed for stochastic hybrid systems. Moreover, large deviation theory does not generate explicit expressions for the prefactor, which we find can contribute significantly to the transition rates.
\medskip

\noindent (iii) We focus on networks operating in the bistable regime, where the simpler QSS diffusion approximation breaks down. There are a growing number of examples of bistability in systems neuroscience, including transitions between cortical up and down states during slow wave sleep \cite{Compte03,Parga07}, working memory \cite{Gutkin01}, and ambiguous perception as exemplified by binocular rivalry \cite{blake11,morenobote07,laing02a,Bressloff12c}. On the other hand, in the case of oscillator networks, a diffusion approximation combined with Floquet theory might be sufficient to capture the effects of noise, including the noise-induced amplification of coherent oscillations or quasicycles \cite{McKane07,Boland08,Bressloff10}. An interesting issue is whether or not the WKB method and matched asymptotics can be applied to a network operating in an excitable regime, where there is a stable low activity resting state such that non-infinitesimal perturbations can induce a large excursion in phase space before returning to the resting state. One of the difficulties with excitable systems is that there does not exist a well-defined separatrix. Nertheless, it is possible to extend the asymptotic methods developed here to the excitable case, as we will show elsewhere within the context of spontaneous action potential generation in a model of an excitable conductance-based neuron with stochastic ion channels.

\section{Stochastic network model and the Chapman-Kolomogorov equation}

Suppose that a network of synaptically coupled spiking neurons is partitioned into a set of $M$ homogeneous populations with $ N$ neurons in each population, $\alpha=1,\ldots,M$. (A straightforward generalization would be take to take each population to consist of ${\mathcal O}(N)$ neurons). Let $\chi$ denote the population function that maps the single neuron index $i=1,\ldots,NM$ to the population index $\alpha$ to which neuron $i$ belongs: $\chi(i)=\alpha$. Furthermore, suppose the synaptic interactions between populations are the same for all neuron pairs. (Relaxing this assumption can lead to additional sources of stochasticity as explored in Ref. \cite{Faugeras09,Touboul11a}). 
Denote the sequence of firing times of the $j$th neuron by $\{T_j^m,\, m \in \Z\}$. The net synaptic current into postsynaptic neuron $i$ due to stimulation by the spike train from presynaptic neuron $j$, with $\chi(i)=\alpha,\chi(j)=\beta$, is taken to have the general form 
$N^{-1}\sum_m\Phi_{\alpha \beta}(t-T_j^m)$, 
where $N^{-1}\Phi_{\alpha\beta}(t)$ represents the temporal filtering effects of synaptic and dendritic processing of inputs from any neuron of population $\beta$ to any neuron of population $\alpha$. For concreteness, we will take exponential synapses so that
\begin{equation}
\label{exps}
\Phi_{\alpha \beta}(t)=w_{\alpha \beta}\Phi(t), \quad\Phi(t)=\tau^{-1}\e^{-t/\tau}H(t)
\end{equation}
(A more general discussion of different choices for $\Phi_{\alpha\beta}(t)$ can be found in the reviews of Ref. \cite{Ermentrout98,Bressloff12}).
Assuming that all synaptic inputs sum linearly, the total synaptic input to the soma of the $i$th neuron, which we denote by $u_i(t)$, is
\begin{eqnarray}
\label{uiu}
u_i(t)&=&\sum_{\beta} \frac{1}{N}\sum_{j;\chi(j)=\beta} \Phi_{\alpha\beta}(t-T_j^m)
=\int_{-\infty}^t\sum_{\beta} \Phi_{\alpha \beta}(t-t')\frac{1}{N}\sum_{j;\chi(j)=\beta}a_j(t')dt'\nonumber\\
\end{eqnarray}
for all $\chi(i)=\alpha$, where 
\begin{equation}
a_j(t)=\sum_{m \in
\Z}\delta(t-T_j^m).
\end{equation} 
That is, $a_j(t)$ represents the output spike train of the $j$th neuron in terms of a sum of Dirac delta functions. (Note that in (\ref{uiu}) we are neglecting any $i$-dependent transients arising from initial conditions, since these decay exponentially for any biophysically based model of the kernels $\Phi_{\alpha \beta}$). In order to obtain a closed set of equations, we have to determine threshold conditions for the firing times $T_i^m$. These take the form
\begin{equation}
\label{Ti}
T_i^m= \inf\{t, t>T_i^{m-1} | V_i(t)=\kappa_{\rm th}, \, \frac{dV_i}{dt}> 0\},
\end{equation}
where $\kappa_{\rm th}$ is the firing threshold and $V_i(t)$ is the somatic membrane potential. The latter is taken to evolve according to a conductance--based model 
\begin{equation}
C  \frac{dV_i}{dt} = -I_{\rm con,i}(V_i,\ldots)+ u_i,
\label{cbi}
\end{equation}
which is supplemented by additional equations for a set of ionic gating variables \cite{Terman}. (The details of the conductance-based model will not be important for the subsequent analysis). Let $a_{\alpha}(t)$ denote the output activity of the $\alpha$th population:
\begin{equation}
\label{A}
a_{\alpha}(t)=\frac{1}{N}\sum_{j;\chi(j)=\alpha}a_j(t),
\end{equation}
and rewrite equation (\ref{uiu}) as
\[
u_i(t)
=\int_{-\infty}^t\sum_{\beta} \Phi_{\alpha \beta}(t-t')a_{\beta}(t')dt',\quad \chi(i)=\alpha.
\]
Since the right-hand side is independent of $i$, it follows that $u_i(t)=u_{\alpha}(t)$ for all $\chi(i)=\alpha$ with
\begin{eqnarray}
u_{\alpha}(t)=\sum_{\beta=1}^M \int_{-\infty}^{t} \Phi_{{\alpha}\beta}(t-t')a_{\beta}(t')dt'.
\label{ust}
\end{eqnarray}
In the case of exponential synapses (\ref{exps}), equation (\ref{ust}) can be converted to the differential equation
\begin{equation}
 \tau \frac{du_{\alpha}}{dt}=-{u_{\alpha}(t)}+\sum_{\beta=1}^M w_{{\alpha}\beta}a_{\beta}(t).
\label{udet}
\end{equation}

In general, equations (\ref{uiu})--(\ref{cbi}) are very difficult to analyze. However, considerable simplification can be obtained if the total synaptic current $u_{i}(t)$ is slowly varying compared to the membrane potential dynamics given by equation (\ref{cbi}). This would occur, for example, if each of the homogeneous subnetworks fired asynchronously \cite{Gerstner02}. 
One is then essentially reinterpreting the population activity variables $u_{\alpha}(t)$ and $a_{\alpha}(t)$ as mean fields of local populations. (Alternatively, a slowly varying synaptic current would occur if the synapses are themselves sufficiently slow \cite{Ermentrout94,Bressloff00}). These simplifying assumptions motivate replacing the output population activity by an instantaneous firing rate $a_{\alpha}(t)=F(u_{\alpha}(t))$ with $F$ identified as the so--called population gain function. 
Equation (\ref{ust}) then forms the closed system of integral equations
\begin{eqnarray}
u_{\alpha}(t) =\int_{-\infty}^{t} \sum_{\beta} \Phi_{\alpha\beta}(t-t')F(u_{\beta}(t'))dt'.
\label{ui2}
\end{eqnarray}
The basic idea is that if neurons in a local population are firing asynchronously then the output activity $a_{\alpha}$ is approximately constant, which means that the synaptic currents are also slowly varying functions of time. A nonlinear relationship between $a_{\alpha}$ and constant input current $u_{\alpha}$ can then be derived using population averaging in order to determine the gain function $F$. One then assumes that the same relationship $a_{\alpha}=F(u_{\alpha})$ also holds for time-dependent synaptic currents, provided that the latter vary slowly with time.
In certain cases $F$ can be calculated explicitly \cite{Abbott93,Gerstner93,Brunel99,Brunel00}. Typically, a simple model of a spiking neuron is used, such as the integrate--and--fire model \cite{Gerstner02}, and the network topology is assumed to be either fully connected or sparsely connected. It can then be shown that under certain conditions, even though individual neurons exhibit Poisson--like statistics, the neurons fire asynchronously so that the total population activity evolves according to a mean--field rate equation with a characteristic gain function $F$. In practice, however, it is sufficient to approximate the firing rate function by a sigmoid:
\begin{equation}
F(u)=\frac{F_0}{1+\e^{-\gamma(u-\kappa)}},
\label{sig}
\end{equation}
where $\gamma,\kappa$ correspond to the gain and threshold respectively.

One of the goals of this paper is to develop a generalization of the neural master equation \cite{Buice07,Buice09,Bressloff09} that incorporates synaptic dynamics. We proceed by taking the ouput activity of a local homogeneous population to be a discrete stochastic variable $A_{\alpha}(t)$ rather than the instantaneous firing rate $a_{\alpha}=F(u_{\alpha})$: 
\begin{equation}
A_{\alpha}(t)=\frac{N_{\alpha}(t)}{N\Delta t},
\label{AN}
\end{equation} 
where $N_{\alpha}(t)$ is the number of neurons in the $\alpha$th population that fired in the time interval $[t-\Delta t,t]$, and $\Delta t$ is the width of a sliding window that counts spikes. The discrete stochastic variables $N_{\alpha}(t)$ are taken to evolve according to a one--step jump Markov process:
\begin{eqnarray}
N_{\alpha}(t)  \rightarrow N_{\alpha}(t)\pm 1:\quad \mbox{transition rate $\Omega_{\pm}(U_{\alpha}(t),N_{\alpha}(t))$},
\label{jump}
\end{eqnarray}
with the synaptic current $U_{\alpha}(t)$ given by (for exponential synapses)
\begin{equation}
 \tau dU_{\alpha}(t)=\left [-{U_{\alpha}(t)}+\sum_{\beta=1}^M w_{{\alpha}\beta}A_{\beta}(t)\right ]dt.
\label{ustoch}
\end{equation}
The transition rates are taken to be (cf. \cite{Bressloff09})
\begin{equation}
\label{Ome}
\Omega_+(u_{\alpha},n_{\alpha})\rightarrow \Omega_+(u_{\alpha})=\frac{N\Delta t}{\tau_{a}}F(u_{\alpha}),\quad \Omega_-(u_{\alpha},n_{\alpha})\rightarrow \Omega_-(n_{\alpha})=\frac{n_{\alpha}}{\tau_{a}}.
\end{equation}
The resulting stochastic process defined by equations (\ref{ustoch}), (\ref{AN}), (\ref{jump}) and (\ref{Ome}) is an example of a stochastic hybrid system based on a piecewise deterministic process. That is,  the transition rate $\Omega_+$ depend on $U_{\alpha}$, with the latter itself coupled to the associated jump Markov according to equation (\ref{ustoch}), which is only defined between jumps, during which $U_{\alpha}(t)$ evolves deterministically. (Stochastic hybrid systems also arise in applications  to genetic networks \cite{Zeisler08,Newby12a} and
to excitable neuronal membranes \cite{Pakdaman10,Buckwar10,Newby11}).  It is important to note that the time constant $\tau_{a}$ cannot be identified directly with membrane or synaptic time constants. Instead, it determines the relaxation rate of a local population to the instantaneous firing rate.

\subsection{Neural master equation}
Previous studies of the neural jump Markov process have effectively taken the limit $\tau \rightarrow 0$ in equation (\ref{ustoch}) so that the continuous variables $U_{\alpha}(t)$ are eliminated by setting $U_{\alpha}(t)=\sum_{\beta}w_{\alpha \beta}A_{\alpha}(t)$. This then leads to a pure birth--death process for the discrete variables $N_{\alpha}(t)$. That is, let $P({\mathbf n},t) = \mbox{Prob}[{\bf N}(t) ={\bf n}]$ denote the probability that the network of interacting populations has configuration ${\mathbf n} = (n_1,n_2,\ldots,n_M)$ at time $t, t >0$, given some initial distribution $P({\mathbf n},0)$ with $0 \leq n_{\alpha}\leq N$. The probability distribution then evolves according to the birth--death master equation \cite{Buice07,Buice09,Bressloff09}
 \begin{eqnarray}
\label{master}
\frac{dP(\n,t)}{dt}= \sum_{\alpha} \left[ ({\mathbb T}_{\alpha}-1) \left (\Omega_{\alpha}^-(\n)P(\n,t) \right ) + ( {\mathbb T}_{\alpha}^{-1}-1)\left (\Omega_{\alpha}^+(\n)P(\n,t)\right )
\right ],
\end{eqnarray}
where 
\begin{equation}
\label{Ome2}
\Omega_{\alpha}^+(\n)=\frac{N\Delta t}{\tau_{a}}F\left (\sum_{\beta}w_{\alpha \beta}n_{\alpha}/N\Delta t \right ),\quad \Omega_{\alpha}^-(\n)=\frac{n_{\alpha}}{\tau_{a}},
\end{equation}
and ${\mathbb T}_{\alpha}$ is a translation operator: ${\mathbb T}_{\alpha}^{\pm 1}F(\n)=F(\n_{\alpha \pm})$ for any function $F$ with $\n_{\alpha \pm}$ denoting the configuration with $n_{\alpha}$ replaced by $n_{\alpha}\pm 1$. Equation (\ref{master}) is supplemented by the boundary conditions $P(\n,t)\equiv 0$ if $n_{\alpha} =N +1$ or $n_{\alpha} =-1$ for some $\alpha$. 
The birth--death master equation (\ref{master}) can be analyzed by adapting various methods from the analysis of chemical master equations including system-size expansions, WKB approximations, and path integral representations \cite{Buice09,Bressloff09,Bressloff10,Buckwar12}. First, suppose that we fix $\Delta t=1$ so that we obtain the Bressloff version of the master equation. Taking the thermodynamic limit $N\rightarrow \infty$ then yields the deterministic activity-based mean--field equation
\begin{equation}
\tau_{\alpha}\frac{dA_{\alpha}}{dt}=-A_{\alpha}(t)+F(\sum_{\beta}w_{\alpha \beta}A_{\alpha}(t)) .
\label{adet}
\end{equation}
(For a detailed discussion of the differences between activity-based and voltage-based neural rate equations, see Refs. \cite{Terman,Bressloff12}).
For large but finite $N$, the master equation (\ref{master}) can be approximated by a Fokker--Planck equation using a Kramers-Moyal or system-size expansion, so that the population activity $A_{\alpha}$ evolves according to a Langevin equation \cite{Bressloff09}. A rigorous probabilistic treatment of the thermodynamic limit of the neural master equation has also been developed \cite{Buckwar12}, extending previous work on chemical master equations \cite{Kurtz71}. Although the diffusion approximation can capture the stochastic dynamics of the neural population at finite times, it can break down in the limit $t\rightarrow \infty$. For example, suppose that the deterministic system (\ref{adet}) has multiple stable fixed points. The diffusion approximation can then account for the effects of fluctuations well within the basin of attraction of a locally stable fixed point. However, there is now a small probability that there is a noise--induced transition to the basin of attraction of another fixed point. Since the probability of such a transition is usually of order $\e^{-cN}$, $c={\mathcal O}(1)$, except close to the boundary of the basin of attraction, such a contribution cannot be analyzed accurately using standard Fokker--Planck methods \cite{vanKampen92}. These exponentially small transitions play a crucial role in allowing the network to approach the unique stationary state (if it exists) in the asymptotic limit $t\rightarrow \infty$, and can be analyzed using a WKB approximation of the master equation together with matched asymptotics \cite{Bressloff10}. In other words, for a multistable neural system, the limits $t\rightarrow \infty$ and $N\rightarrow \infty$ do not commute, as previously noted for chemical systems \cite{Hanggi84}.

Now suppose that we take the limit $N\rightarrow \infty,\Delta t \rightarrow 0$ such that $N\Delta t=1$. We then recover the neural master equation of Buice {\em et. al.} \cite{Buice07,Buice09}. In this case there is no small parameter that allows us to construct a Langevin approximation to the master equation. Nevertheless, it is possible to determine the moment hierarchy of the master equation using path integral methods or factorial moments, based on the observation that the network operates in a Poisson-like regime. The role of the sliding window size $\Delta t$ is crucial in understanding the difference between the two versions of the master equation. First, it should be emphasized that the stochastic models are keeping track of {\em changes} in population spiking activity. If the network is operating close to an asynchronous state for large $N$, then one-step changes in population activity could occur relatively slowly so there is no need to take the limit $\Delta t \rightarrow 0$. On the other hand, if population activity is characterized by a Poisson process then it is necessary to take the limit $\Delta t \rightarrow 0$ in order to maintain a one-step process. However, given the existence of an arbitrarily small time-scale $\Delta t$, it is no longer clear that one is justified in ignoring synaptic dynamics by taking the limit $\tau\rightarrow 0$ in equation (\ref{ustoch}). This observation motivates the approach taken in this paper, in which we incorporate synaptic dynamics into the neural master equation. In the following, we will assume that the network operates in the Poisson-like regime in the absence of synaptic dynamics.

\subsection{Neural Chapman-Kolmogorov equation}
Let us now return to the full stochastic hybrid system.
Denote the random state of the full model at time $t$ by the vector ${\bf X}(t)= \{(U_{\alpha}(t),N_{\alpha}(t)); \alpha = 1,\ldots,M\}$. Introduce the corresponding probability density 
\begin{equation}
\mbox{Prob}\{U_{\alpha}(t)\in (u_{\alpha},u_{\alpha}+du, N_{\alpha}(t)=n_{\alpha};\alpha=1,\ldots,M\}=p(\u,\n,t|\u_0,\n_0,0)d\u,
\end{equation}
with $\u=(u_1,\ldots,u_M)$ and $\x=(\u,\n)$. It follows from equations (\ref{ustoch}), (\ref{AN}), (\ref{jump}) and (\ref{Ome}) that the probability density evolves according to the Chapman-Kolmogorov equation
\begin{eqnarray}
\label{CK}
&&\frac{\partial p}{\partial t}+\frac{1}{\tau}\sum_{\alpha}\frac{\partial [v_{\alpha}({\bf x})p(\bf{x},t)]}{\partial u_{\alpha}}
\\&&\qquad =\frac{1}{\tau_a}\sum_{\alpha} \left[ ({\mathbb T}_{\alpha}-1) \left (\omega_-(n_{\alpha})p({\bf x},t) \right ) + ( {\mathbb T}_{\alpha}^{-1}-1)\left (\omega_+(u_{\alpha})p(\bf{x},t)\right )\right ],\nonumber 
\end{eqnarray}
with
\begin{equation}
\omega_+(u_{\alpha})=F(u_{\alpha}),\quad \omega_-(n_{\alpha})=n_{\alpha},\quad v_{\alpha}(\x)=-u_{\alpha}+\sum_{\beta}w_{\alpha \beta}n_{\beta}.
\end{equation}
We have taken the limit $N\rightarrow \infty$, $\Delta t \rightarrow 0$ with $N\Delta t=1$. Note that equation (\ref{CK}) can be expressed in the general form of equation (\ref{0}). Thus, in the limit $\tau\rightarrow 0$ we recover the master equation of Buice {\em et. al.} \cite{Buice07,Buice09}, whereas in the limit $\tau_a\rightarrow 0$ we obtain the mean-field equations
\begin{eqnarray}
 \tau \frac{du_{\alpha}}{dt}&=&\langle v_{\alpha}\rangle(\u(t)) \equiv \sum_{\n}v_{\alpha}(\u(t),\n)\rho(\u(t),\n)\nonumber\\
 &=& -{u_{\alpha}(t)}+\sum_{\beta=1}^M w_{{\alpha}\beta}\sum_{\n} n_{\beta}\rho(\u(t),\n).
\label{udet2}
\end{eqnarray}
It can be shown that $\rho(\u,\n)$ is given by a compound Poisson process with rates $F(u_{\alpha})$, consistent with the operating regime of the Buice {\em et. al.} master equation \cite{Buice07,Buice09}. Hence, in this limit,
\begin{equation}
\langle n_{\beta}\rangle = F(u_{\beta}),
\end{equation}
and (\ref{udet2}) reduces to the standard voltage or current-based activity equation.

\subsection{Quasi-steady-state (QSS) diffusion approximation}

In this paper, we will consider the regime in which the transitions between different firing states are much faster than the synaptic dynamics. Hence, fixing the units of time by setting $\tau=1$, we take $\tau_{\alpha}/\tau =\epsilon \ll 1$. Since $\epsilon \ll 1$, there will typically be a large number of transitions between different firing states $\n$ while the synaptic currents $\u$ hardly change at all. This suggests that the system will rapidly converge to the steady-state $\rho(\u,\n)$ (if it exists) given by equation (\ref{1}). The full probability density will then be perturbed away from this steady-state density as $\u$ slowly evolves. However, if $\epsilon \ll 1$ then these perturbations will be small and the solution will tend to remain close to the steady state. The resulting perturbations can then be analyzed using a quasi-steady-state (QSS) diffusion or adiabatic approximation, in which the CK equation reduces to a Fokker--Planck (FP) equation. This method was first developed from a probabilistic perspective by Papanicolaou \cite{Papanicolaou}, see also \cite{Gardiner09}. It has subsequently been applied to a wide range of problems in biology, including cell movement \cite{Othmer88,Hillen00}, traveling-wavelike behavior in models of slow axonal transport \cite{reed1990,Friedman05,Friedman07}, and molecular motor-based models of random intermittent search \cite{Newby09,Newby10a,Newby10b,Bressloff11}.

Consider a Chapman Kolmogrov equation of the general form (see equation (\ref{0}))
\begin{equation}
\label{A0}
\frac{\partial p}{\partial t}=-\sum_{{\alpha}=1}^M\frac{\partial}{\partial u_{\alpha}}(v_{\alpha}(\u,\n)p(\u,\n,t))+\frac{1}{\epsilon}\sum_{\m}A(\n,\m;\u)p(\u,\m,t),
\end{equation}
with $\epsilon \ll 1$. We assume that for fixed $\u$, the tensor $A(\n,\m;\u)$ is equivalent to a transition matrix. That is, suppose we relabel the discrete states according to $\n \rightarrow I,\m\rightarrow J$ with $I,J=0,1,\ldots ,\chi$ for $\chi=(N+1)^M$ and set $A(\n,\m;\u)=A_{IJ}(\u)$, $p(\u,\n,t)=p_I(\u,t)$. Then the $\chi \times\chi $ matrix ${\bf A}(\u)$ with elements $A_{IJ}(\u)$ is taken to be irreducible and to have a simple zero eigenvalue with corresponding left eigenvector ${\bf 1}$ whose components are all unity. In other words, $\sum_IA_{IJ}(\u)=0$ for all $J$. The Perron-Frobenius Theorem then ensures that all other eigenvalues are negative and the continuous-time Markov process for fixed $\u$,
\[\frac{d p_I}{d t}=\frac{1}{\epsilon}\sum_{J}A_{IJ}p_J(\u,t),\]
has a globally attracting steady-state $\rho_I(\u)$ such that $p_I(\u,t)\rightarrow \rho_I(\u)$ as $t\rightarrow \infty$. Here $\rho_I(\u)$ is the unique right eigenvector corresponding to the zero eigenvalue of ${\bf A}(\u)$, that is, $\sum_JA_{IJ}(\u)\rho_J(\u)=0$. In terms of the original notation, we have 
\begin{equation}
\label{1}
\sum_{\n}A(\n,\m;\u)=0,\quad \sum_{\m}A(\n,\m;\u)\rho(\u,\m)=0. 
\end{equation}
In the following it will be convenient to introduce the summation operator
\begin{equation}
[{\mathbf 1}^T\circ f](\u)=\sum_{\n}f(\u,\n)
\end{equation}
for any function $f(\n,\u)$.

The first step in the QSS reduction is to decompose the probability density as
\begin{equation}
\label{C}
p(\u,\n,t)=C(\u,t)\rho(\u,\n)+\epsilon w(\u,\n,t),
\end{equation}
where $\rho(\u,\n)$ is the steady-state density given by equation (\ref{1}), and
${\mathbf 1}^T\circ p =C$, ${\mathbf 1}^T\circ w=0$. Applying the summation operator to both sides of equation (\ref{A0}) and using $\sum_{\n} A(\n,\m,\u)=0$ gives
\begin{equation}
\label{CC}
\frac{\partial C}{\partial t}=-{\mathbf 1}^T\circ\left [\sum_{{\alpha}=1}^M\frac{\partial}{\partial u_{\alpha}}(v_{\alpha}[C\rho+\epsilon w])\right ].
\end{equation}
Next substitute (\ref{C}) into equation (\ref{A0}) to give
\begin{equation*}
\frac{\partial C}{\partial t}\rho +\epsilon \frac{\partial w}{\partial t}={\bf A}\circ w-\sum_{{\alpha}=1}^M\frac{\partial}{\partial u_{\alpha}}(v_{\alpha}[C\rho+\epsilon w]),
\end{equation*}
where
\begin{equation*}
[{\bf A}\circ w](\u,\n,t)=\sum_{\m}A(\n,\m;\u)w(\u,\m,t).
\end{equation*}
Combining with equation (\ref{CC}) shows that
\begin{equation*}
\epsilon \frac{\partial w}{\partial t}={\bf A}\circ w+\rho {\mathbf 1}^T\circ\left [\sum_{{\alpha}=1}^M\frac{\partial}{\partial u_{\alpha}}(v_{\alpha}[C\rho+\epsilon w])\right ]-\sum_{{\alpha}=1}^M \frac{\partial}{\partial u_{\alpha}}(v_{\alpha}[C\rho+\epsilon w]).
\end{equation*}
Collecting terms of leading order in $\epsilon$ yields
\begin{equation}
{\bf A}\circ w=\sum_{{\alpha}=1}^M \frac{\partial}{\partial u_{\alpha}}(v_{\alpha}C\rho)-\rho {\mathbf 1}^T\circ\left [\sum_{{\alpha}=1}^M\frac{\partial}{\partial u_{\alpha}}(v_{\alpha}C\rho)\right ].
\end{equation}
The Fredholm Alternative Theorem \cite{Ramm01} ensures that this equation has a unique solution for $w$ subject to the constraint ${\mathbf 1}^T\circ w=0$; we formally denote this solution as
\begin{equation}
w\sim {\bf A}^{\dagger}\circ \sum_{{\alpha}=1}^M \frac{\partial}{\partial u_{\alpha}}(v_{\alpha}C\rho)- ({\bf A}^{\dagger}\circ\rho) {\mathbf 1}^T\circ\left [\sum_{{\alpha}=1}^M\frac{\partial}{\partial u_{\alpha}}(v_{\alpha}C\rho)\right ],
\end{equation}
where ${\mathbf A}^{\dagger}$ is the pseudoinverse operator. Substituting for $w$ back into equation (\ref{CC}) and ignoring ${\mathcal O}(\epsilon^2)$ terms finally gives the Fokker--Planck equation
\begin{equation}
\label{FPC}
\frac{\partial C}{\partial t}=-\sum_{\alpha=1}^M \frac{\partial}{\partial u_{\alpha}}(V_{\alpha}C)+\epsilon \sum_{\alpha=1}^M\sum_{\beta=1}^M\frac{\partial}{\partial u_{\alpha}}\left (D_{\alpha\beta}\frac{\partial C}{\partial u_{\beta}}\right ),
\end{equation}
where
\begin{equation}
V_{\alpha}(\u)=\sum_{\n}v_{\alpha}(\u,\n)\rho(\u,\n)
\end{equation}
and
\begin{equation}
D_{\alpha \beta}=\sum_{\m,\n}[v_{\alpha}(\m,\u)-V_{\alpha}(\u)]A^{\dagger}(\m,\n;\u)[v_{\beta}(\u,\n)-V_{\beta}(\u)]\rho(\u,\n).
\end{equation}
Note that we have expressed $D_{\alpha\beta}$ in a more symmetric form using the fact that $\sum_{\m}A(\m,\n,\u)=0$.
Finally, let us introduce the function $z_{\alpha}(\u,\m)$, which satisfies the equation
\begin{equation}
\label{zz}
\sum_{\n}z_{\alpha}(\u,\n)A(\n,\m;\u)=-[v_{\alpha}(\u,\m)-V_{\alpha}(\u)].
\end{equation}
Since $\sum_{\m}\rho(\u,\m)[v_{\alpha}(\u,\m)-V_{\alpha}(\u)]=0$, it follows from the Fredholm alternative that
\begin{equation}
z_{\alpha}(\u,\n)=-\sum A^{\dagger}(\m,\n;\u)[v_{\alpha}(\u,\m)-V_{\alpha}(\u)]
\end{equation}
and thus
\begin{equation}
\label{DD}
D_{\alpha \beta}=\sum_{\n}z_{\alpha}(\u,\n)[v_{\beta}(\u,\n)-V_{\beta}(\u)]\rho(\u,\n).
\end{equation}
Hence, under the QSS approximation, the stochastic dynamics is characterized by Gaussian fluctuations about the mean-field equations (\ref{udet2}). However, as in the case of the system-size expansion of the neural master equation (\ref{master}) for large $N$ and $\Delta t=1$, approximating the Chapman-Kolmogorov equation (\ref{CK}) by a Fokker-Planck equation for small $\epsilon$ breaks down when considering rare event transitions between metastable states. This particular issue has recently been addressed within the context of stochastic ion channels and Hodgkin-Huxley dynamics \cite{Newby11}, as well as gene networks \cite{Newby12a}, using asymptotic methods developed in \cite{Keener11,Newby12}. In the following sections, we will extend such methods to the neural CK equation.

\section{Metastable states in a one-population model}

In order to develop the basic analytical framework, consider the simple case of a single recurrent population ($M=1$) evolving according to the CK equation
\begin{eqnarray}
\label{CK1}
&&\frac{\partial p}{\partial t}+\frac{\partial [v(u,n)p(u,n,t)]}{\partial u}
\\&&\qquad =\frac{1}{\epsilon} [\omega_+(u)p(u,n-1,t)+\omega_-(n+1)p(u,n+1,t)\nonumber 
\\&& \qquad \qquad -(\omega_+(u)+\omega_-(n))p(u,n,t) ],\nonumber 
\end{eqnarray}
 with boundary condition $p(u,-1,t)\equiv 0$, drift term
 \begin{equation}
 v(u,n)=-u+wn,
 \end{equation}
 and transition rates
 \begin{equation}
 \label{rates1}
\omega_+(u) =F(u),\quad \omega_-(n)=n.
\end{equation}
Following the general discussion in \S 2.3, we expect the finite-time behavior of the stochastic population to be characterized by small perturbations about the stable steady--state of the pure birth--death process
\begin{eqnarray}
\label{BD1}
&&\frac{\partial p}{\partial t} =\frac{1}{\epsilon} [\omega_+(u)p(u,n-1,t)+\omega_-(n+1)p(u,n+1,t)\nonumber 
\\&& \qquad \qquad -(\omega_+(u)+\omega_-(n))p(u,n,t) ],\nonumber 
\end{eqnarray}
with $u$ treated as a constant over time-scales comparable to the relaxation time of the birth-death process. The equation for the steady--state distribution $\rho(u,n)$ can be written as \cite{Gardiner09}
\begin{equation*}
0=J(u,n+1)-J(u,n),
\end{equation*}
with $J(u,n)$ the probability current,
\begin{equation*}
J(u,n)=\omega_-(n)\rho(u,n)-\omega_+(u)\rho(u,n-1) .
\end{equation*}
Since $\omega_-(0)=0$ and $\rho(u,-1)=0$, it follows that $J(u,0)=0$ and $J(u,n)=0$ for all $n\geq 0$. 
Hence,
\begin{equation}
\label{ss}
\rho(u,n)=\frac{\omega_+(u)}{\omega_-(n)}\rho(u,n-1)=\rho(u,0)\prod_{m=1}^n\frac{\omega_+(u)}{\omega_-(m)}
\end{equation}
with $\rho(u,0)=1-\sum_{m=1}^{\infty}\rho(u,m)$. 

\begin{figure}[htbp]
\begin{center}
\includegraphics[width=8cm]{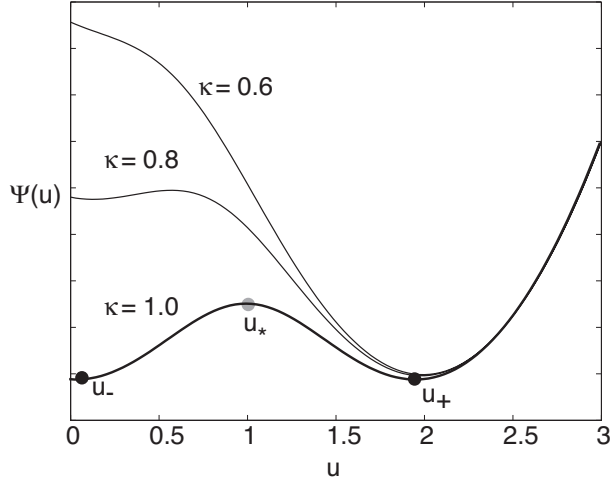}
\caption{\small Bistable potential $\Psi$ for the deterministic network satisfying $\dot{u}=-u+F(u)=-d\Psi/du$ with $F$ given by the sigmoid (\ref{sig}) for $\gamma = 4$, $\kappa=1.0$, $F_0=2$. There exist two stable fixed points $u_{\pm}$ separated by an unstable fixed point $u_*$. As the threshold $\kappa$ is reduced the network switches to a monostable regime via a saddle-node bifurcation.}
\label{fig1}
\end{center}
\end{figure}

Substituting the explicit expressions (\ref{rates1}) for the transition rates, we have
\begin{equation}
\label{ss1}
\rho(u,n)=\rho(u,0)\prod_{m=1}^n\frac{ F(u)}{m}=\rho(u,0)\frac{(F(u))^n}{n!}.
\end{equation}
It follows that
\begin{equation}
\rho(u,0)=\frac{1}{1+\sum_{n=1}^{\infty}(F(u))^n/{n!}}=\e^{-F(u)}.
\end{equation}
so the steady-state density is given by a Poisson process,
\begin{equation}
\label{poiss}
\rho(u,n)=\frac{[F(u)]^n\e^{-F(u)}}{n!}.
\end{equation}
The mean number of spikes is thus $\langle n\rangle =F(u)$,
and the mean-field equation obtained in the $\epsilon \rightarrow 0$ limit is
\begin{equation}
\label{rate}
\frac{du}{dt}=-u+wF(u)\equiv -\frac{d\Psi}{du}.
\end{equation}
The sigmoid function $F(u)$ given by (\ref{sig}) is a bounded, monotonically increasing function of $u$ with $F(u)\rightarrow F_0$ as $u\rightarrow \infty$ and $F(u)\rightarrow 0$ as $u\rightarrow -\infty$. Moreover, $F'(u)=\gamma F_0/[4\cosh^2(\gamma(u-\kappa)/2)]$ so that $F(u)$ has a maximum slope at $u=\kappa$ given by $\gamma F_0/4$. It follows that the function $-u+wF(u)$ only has one zero if $w\gamma F_0< 4$ and this corresponds to a stable fixed point. On the other hand, if $w\gamma F_0 >4$ then, for a range of values of the threshold $\kappa$,  $[\kappa_1,\kappa_2]$, there exists a pair of stable fixed points $u_{\pm}$ separated by an unstable fixed point $u_*$ (bistability). A stable/unstable pair vanishes via a saddle-node bifurcation at $\kappa=\kappa_1$ and $\kappa=\kappa_2$. This can also be seen graphically by plotting the potential function $\Psi(u)$, whose minima and maxima correspond to stable and unstable fixed points of the mean-field equation. An example of the bistable case is shown in  Fig. \ref{fig1}.

The problem we wish to address is how to analyze the effects of fluctuations (for $0< \epsilon \ll 1$) on rare transitions between the metastable states $u_{\pm}$ of the underlying mean-field equation. As highlighted in \S 2.3, it is not possible to use a QSS diffusion approximation, since this only captures finite-time fluctuations within the basin of attraction of a given metastable state. Therefore, we will proceed using the asymptotic methods recently introduced to analyze spontaneous action potentials in conductance-based single neuron model with stochastic ion channels \cite{Newby11}. As a first step, it is convenient to introduce the vector-valued probability density $\p(u,t)=(p_0(u,t),\ldots,p_n(u,t),\dots)$ with $p_n(u,t)=p(u,n,t)$. Equation (\ref{CK}) can then be rewritten in the matrix form\footnote{Recall that we have taken the limit $N\rightarrow \infty, \Delta t\rightarrow 0$ such that $N\Delta t=1$. Hence, we are dealing with infinite matrices. However, this does not cause any problems. Indeed, we could equally well proceed by assuming that $N$ is finite and $\Delta t =1/ N$, performing all calculations, and then taking the limit $N\rightarrow \infty$. The advantage of working with infinite $N$ is that the steady-state density $\rho$ is given by a Poisson process and we don't have to worry about boundary conditions at $n=N$.} 

\begin{equation}
\label{CKv}
\frac{\partial \p}{\partial t}=-\frac{\partial}{\partial u}({\bf V}(u)\p)+\frac{1}{\epsilon}{\bf A}{\bf p}.
\end{equation}
For given $u$, the diagonal drift matrix ${\bf V}$ has non-zero entries
\begin{equation}
\label{Vcomp}
V_{n,n}(u)=v_n(u)\equiv v(u,n)=-u+w n,
\end{equation}
and the tridiagonal transition matrix ${\bf A}$ has entries
\begin{equation}
A_{n,n-1}=\omega_+(u),\quad A_{n,n} =-\omega_+(u)-\omega_-(n),\quad A_{n,n+1}=\omega_-(n+1).
\label{Acomp}
\end{equation}
Since ${\bf A}$ is a transition matrix, its columns sum to zero, it has one zero eigenvalue, and all other eigenvalues are negative. In particular, 
\begin{equation}
\label{null}
{\bf 1}^T {\bf A}=0,\quad {\bf A}{\bm \rho}=0
\end{equation}
with ${\bf 1}^T=(1,1,\ldots)$ and $\rho_n=\rho(u,n)$, where $\rho(u,n)$ is the steady state density (\ref{ss1}). In matrix notation, the mean-field equation (\ref{rate}) can be written as
\begin{equation}
\label{ud}
\frac{du}{dt}= {\bf v}(u)\cdot {\bm \rho}(u).
\end{equation}
Finally, note that since $v_n(u)=-u+wn$, it follows that if the initial synaptic current $u_0>0$ then $u(t)>0$ for all $t>0$. 
\subsection{Quasistationary approximation}
Suppose that the neural population starts in the left--hand well of the potential function $\Psi(u)$, see Fig. \ref{fig1}, at the stable low activity state $u_-$. On short time scales the solution rapidly converges to a quasi stationary solution that is only distributed across the left well. However, on a longer time scale, probability slowly leaks into the right well until the full stationary solution is reached. In order to estimate the exponentially small transition rate from the left to right well, we place an absorbing boundary at the unstable fixed point $u_*$. (The subsequent time to travel from $u_*$ to the high activity fixed point $u_+$ is insignificant, and can be neglected). Thus, the CK equation (\ref{CKv}) is supplemented by the absorbing boundary conditions
\begin{equation}
p_n(u_*,t)=0, \, \mbox{for} \, n=0,\ldots,k-1,
\end{equation}
where $0<k<\infty$ is the number of firing states for which the drift $v(u_*,n)<0$. 
The initial condition is taken to be
\begin{equation}
\label{icon}
p_n(u,0)=\delta(u-u_-)\delta_{n,n_0}.
\end{equation}
Let $T$ denote the (stochastic) first passage time for which the system first reaches $u_*$, given that it started at $u_-$. The distribution of first passage times is related to the survival probability that the system hasn't yet reached $u_*$:
\begin{equation}
S(t)\equiv \sum_{n=0}^{\infty}\int_0^{u_{*}} p_n(u,t)du .
\end{equation}
That is, $\mbox {Prob}\{t>T\} =S(t)$ and the first passage time density is
\begin{equation}
f(t)=-\frac{dS}{dt}=-\sum_{n=0}^{\infty}\int_0^{u_{*}} \frac{\partial p_n}{\partial t}(u,t)du .
\end{equation}
Substituting for $\partial p_n/\partial t$ using the CK equation (\ref{CKv}) shows that
\begin{eqnarray}
f(t)=\sum_{n=0}^{\infty}\int_0^{u_{*}}  \frac{\partial [v(u,n) p_n(u,t)]}{\partial u}du=\sum_{n=0}^{\infty}v(u_*,n)p_n(u_*,t).
\label{fTP}
\end{eqnarray}
We have used the fact that ${\bf 1}^T{\bf A}=0$ and $\p(0,t)=0$. The first passage time density can thus be interpreted as the probability flux $J(u_*,t)$ at the absorbing boundary, since we have the conservation law
\begin{equation}
\sum_{n=0}^{\infty}\frac{\partial p_n}{\partial t}=-\frac{\partial J}{\partial u},\quad J(u,t)=\sum_{n=0}^{\infty}v(u,n)p_n(u,t).
\end{equation}

The probability flux at the absorbing boundary can be approximated using a spectral projection method \cite{Ward98, hinch05a, Keener11, Newby11}.
Consider an eigenfunction expansion of the form
\begin{equation}
\p(u,t)=\sum_{j=0}^{\infty}C_j\e^{-\lambda_jt}{\bm \phi}_j(u),
\end{equation}
where the eigenfunctions satisfy the equation
\begin{equation}
\label{Lhat}
\widehat{L}{\bm \phi}_j\equiv \frac{d}{du}({\bf V}{\bm \phi}_j)-\frac{1}{\epsilon}{\bf A}{\bm \phi}_j=\lambda_j{\bm \phi}_j,
\end{equation}
together with the boundary conditions
\begin{equation}
({\bm \phi}_j)_n(u_*)=0,\, \mbox{for}\, n=0,\ldots,k-1 .
\end{equation}
If the absorbing boundary is replaced by a reflecting boundary, then there is a single zero eigenvalue $\lambda_0$, for which the resulting stationary density ${\bf p}_s$ is the corresponding eigenfunction. On the other hand, when there is an absorbing boundary, the stationary solution ceases to exist due to a nonzero probability flux through $u_*$. Moreover, $\lambda_0$ is perturbed away from zero but is exponentially small compared to the remaining eigenvalues. In other words, $\lambda_0={\mathcal O}(\e^{-L/\epsilon})$ for some $L >0$ and $\lambda_j={\mathcal O}(1)$, $j \geq 1$. It follows that all other eigenmodes decay to zero much faster than the perturbed stationary density. Thus, at large times we have the quasistationary approximation
\begin{equation}
\label{qa}
\p(u,t)\sim C_0 \e^{-\lambda_0 t}{\bm \phi}_0(u) .
\end{equation}
Substituting such an approximation into equation (\ref{fTP}) implies that
\begin{equation}
f(t)\sim {\bf v}(u_*)\cdot {\bm \phi}_0(u_*)C_0 \e^{-\lambda_0 t},\quad \lambda_1t\gg 1,
\label{FT}
\end{equation}

The next step of the spectral projection method is to define a set of eigenfunctions for the adjoint operator, which satisfy the equation
\begin{equation}
\label{adj}
\widehat{L}^*{\bm \xi}_j\equiv -{\bf V} \frac{d}{du}({\bm \xi}_j)-\frac{1}{\epsilon}{\bf A}^T{\bm \xi}_j=\lambda_j{\bm \xi}_j,
\end{equation}
and the boundary conditions
\begin{equation}
({\bm \xi}_j)_n(u_*)=0,\quad n\geq k.
\end{equation}
The two sets of eigenfunctions form a biorthogonal set with respect to the underlying inner product, which is taken to be
\begin{equation}
\langle {\bf f},{\bf g}\rangle =\int_{0}^{u_*}{\bf f}^T(u){\bf g}(u)du.
\end{equation}
Now consider the identity
\begin{equation}
\label{id}
\langle {\bm \phi}_0,\widehat{L}^* {\bm \xi}_0\rangle = \lambda_0 \langle {\bm \phi}_0,{\bm \xi}_0\rangle.
\end{equation}
Suppose that the exact eigenfunction ${\bm \phi}_0$ satisfying the absorbing boundary conditions can be approximated by a quasistationary solution ${\bm \phi}_{\epsilon}$ for which $\widehat{L} {\bm \phi}_{\epsilon}=0$ without any absorbing boundaries. Under such an approximation, integrating by parts the left-hand side of equation (\ref{id}) picks up a boundary term so that
\begin{equation}
\label{l0}
 \lambda_0\sim - \frac{{\bm \phi}_{\epsilon}^T(u_*){\bf V}(u_*) {\bm \xi}_0(u_*)}{ \langle {\bm \phi}_{\epsilon},{\bm \xi}_0\rangle}.
\end{equation}
The calculation of the principal eigenvalue $\lambda_0$ thus reduces to the problem of determining the quasistationary density ${\bm \phi}_{\epsilon}$ and the exact adjoint eigenfunction $\xi_0$ using perturbation methods (see below). Once $\lambda_0$ has been evaluated, we can then identify the mean first passage time $\langle T\rangle $ with $\lambda_0^{-1}$. In order to establish this, we derive an alternative approximation for $\lambda_0$ by starting from the identity $\langle \widehat{L}{\bm \phi}_0, {\bm \xi}_0\rangle =\lambda_0\langle {\bm \phi}_0,{\bm \xi}_0\rangle$ and making the approximation ${\bm \xi}_0(u)\sim {\bf 1}$, which is valid outside a boundary layer around the absorbing boundary. Integration by parts now yields
\begin{equation}
\label{l1}
 \lambda_0\sim \frac{{\bm \phi}_{0}^T(u_*){\bf V}(u_*){\bf 1}}{ \langle {\bm \phi}_{0},{\bf 1}\rangle} \sim \frac{{\bf v}(u_*)\cdot {\bm \phi}_0(u_*)}{ \langle {\bm \phi}_{0},{\bf 1}\rangle}.
\end{equation}
Moroever, from the initial condition (\ref{icon}) and the quasistationary approximation (\ref{qa}), we have
\[ \langle {\bm \xi}_0, \p(0)\rangle = ({\bm \xi}_0)_{n_0}(u_-) \sim C_0 \langle {\bm \xi}_0, {\bm \phi}_0\rangle .\]
so that for ${\bm \xi}_0(u)\sim {\bf 1}$, 
\begin{equation}
C_0\sim \frac{1}{\langle {\bf 1}, {\bm \phi}_0\rangle}.
\end{equation}
Equation (\ref{FT}) then shows that the first passage time density reduces to
\begin{equation}
f(t)\sim \lambda_0 \e^{-\lambda_0 t}
\end{equation}
and $\langle T\rangle =\int_0^{\infty}tf(t)dt \sim 1/\lambda_0$.

\subsection{WKB method and the quasi-stationary density}

We now use the Wentzel-Kramers-Brillouin (WKB) method \cite{Hanggi84, Matkowsky90, Dykman94, Stein97, Schuss10} to compute an approximation ${\bm \phi}_{\epsilon}$ of ${\bm \phi}_0$ that does not satisfy the absorbing boundary condition. (Although such methods have been applied extensively to Fokker-Planck equations and master equations, the extension to CK equations of the form (\ref{CKv}) is relatively recent). We thus seek an approximate solution of $\widehat{L}{\bm \phi}_{\epsilon}=0$ of the WKB form
\begin{equation}
\label{WKB}
\vphi(u)={\bf R}(u)\exp\left (-\frac{\Phi(u)}{\epsilon}\right ),
\end{equation}
where $\Phi(u)$ is a scalar potential. Substituting into equation (\ref{Lhat}) gives
\begin{equation}
\label{AAA}
({\bf A}+\Phi'{\bf V}){\bf R}=\epsilon ({\bf V}{\bf r})'+\lambda_0{\bf R},
\end{equation}
where $'$ denotes $d/du$. Introducing the asymptotic expansions ${\bf R}\sim {\bf R}_0+\epsilon {\bf R}_1$ and $ \Phi=\Phi_0+\epsilon \Phi_1$, and using the fact that $\lambda_0={\mathcal O}(\e^{-L/\epsilon})$, the leading order equation is
\begin{equation}
{\bf A}{\bf R}_0=-\Phi_0'{\bf V}{\bf R}_0.
\end{equation}
The diagonal components of ${\bf V}(u)$ are invertible almost everywhere for $u>0$. Thus we can identify $-\Phi_0'$ and ${\bf R}_0$ as an eigenpair of the eigenvalue problem
\begin{equation}
\label{eigM}
{\bf M}{\bm \psi}=\mu {\bm \psi},\quad {\bf M}\equiv{\bf V}^{-1}{\bf A}.
\end{equation}
Positivity of the probability density $\vphi$ requires positivity of the corresponding eigenfunction ${\bm \psi}$. For fixed $u$, the matrix ${\bf M}$ has ${\bf \rho}$ as a right null vector and ${\bf v}$ as a left nullvector. These results follow from ${\bf A}{\rho}=0$ and ${\bf v}^T{\bf V}^{-1}{\bf A}={\bf 1}^T{\bf A}=0$. Thus one positive eigenfunction is ${\bm \psi}_0=\rho$ with corresponding eigenvalue $\mu_0=0$. However, such a solution is not admissible since $\Phi_0'=0$ and $\Phi_0={\rm constant}$. Since $v_n(u)$ for fixed $0<u$ changes sign as $n$ increases from zero, theorem 3.1 of \cite{Keener11} ensures that there exists one other positive eigenfunction, which we denote by ${\bm \psi}_1$, whose corresponding eigenvalue $\mu_1$ varies with $u$ in such a way that the corresponding WKB approximation is valid. Here we will construct such an eigenfunction explicitly.

Using the explicit expressions for ${\bf V}$ and ${\bf A}$, equations (\ref{Vcomp}) and (\ref{Acomp}), the eigenvalue equation can be written in component form as
\begin{equation}
 F(u)\psi_{n-1}-( F(u)+n)\psi_n+(n+1)\psi_{n+1}=\mu (-u+wn)\psi_n
\end{equation}
Trying a solution for ${\bm \psi}_1$ of the form
\begin{equation}
({\bm \psi})_n=\frac{\Lambda^n}{n!},
\end{equation}
yields the following equation relating $\Lambda$ and the corresponding eigenvalue $\mu_1$:
\begin{equation*}
\left [\frac{ F(u)}{\Lambda}-1\right ]n+\Lambda-F(u)=\mu_1 (-u+wn).
\end{equation*}
We now collect terms independent of $n$ and linear in $n$ to obtain the pair of equations
\begin{equation*}
\mu_1=\frac{1}{w}\left [\frac{F(u)}{\Lambda}-1\right ],\quad \Lambda = F(u)-\mu u.
\end{equation*}
We hence deduce that
\begin{equation}
\label{mu1}
\Lambda=\frac{u}{w},\quad \mu_1=\frac{1}{w}\left [\frac{wF(u)}{u}-1\right ],
\end{equation}
and
\begin{equation}
\label{psi1}
({\bm \psi}_1)_n={\mathcal M}\frac{1}{n!}\left (\frac{u}{w}\right )^n,
\end{equation}
where ${\mathcal M}$ is a normalization factor.
Note that $\mu_1(u)$ vanishes at the fixed points $u_-,u_*$ of the mean-field equation (\ref{rate}) with $\mu_1(u)>0$ for $0<u<u_-$ and $\mu_1(u)<0 $ for $u_-<u<u_*$. Moreover, comparing equation (\ref{ss1}) with (\ref{psi1}) establishes that ${\bm \psi}_1(u)=\rho(u)$ at the fixed points $u_*,u_{\pm}$. In conclusion ${\bf R}_0={\bm \psi}_1$ and the effective potential $\Phi_0$ is given by
\begin{equation}
\label{pot}
\Phi_0(u)=-\int_{u_-}^u \mu_1(y)dy.
\end{equation}
The effective potential is defined up to an arbitrary constant, which has been fixed by setting $\Phi_0(u_-)=0$.

Proceeding to the next order in the asymptotic expansion of equation (\ref{AAA}), we have
\begin{equation}
({\bf A}+\Phi_0'{\bf V}){\bf R}_1=({\bf V}{\bf R}_0)'-\Phi_1'{\bf V}{\bf R}_0.
\end{equation}
Since ${\bf R}_0={\bm \psi}_1$ and $\Phi_0'=-\mu_1$, it follows from the Fredholm alternative that
\begin{equation}
\label{bat}
\Phi_1'=\frac{{\bm \eta}^T({\bf V}{\bm \psi}_1)'}{{\bm \eta}^T{\bf V}{\bm \psi}_1},
\end{equation}
where ${\bm \eta}$ is the left null vector of ${\bf A}-\mu_1{\bf V}$. Using equations (\ref{Vcomp}) and (\ref{Acomp}), the components of ${\bm \eta}$ satisfy the explicit equation
\begin{equation}
 F(u) \eta_{m+1}-(F(u) +m)\eta_m+m\eta_{m-1}=\mu_1 [-u+m w]\eta_m .
\end{equation}
Trying a solution of the form $\eta_m=\Gamma^m$ yields
\begin{equation}
(F(u))\Gamma -( F(u) +m)+m\Gamma^{-1}=\mu_1 [-u+m w].
\end{equation}
$\Gamma$ is then determined by canceling terms linear in $m$, which finally gives
\begin{equation}
{\bm \eta}_n=\left (\frac{u}{wF(u)}\right )^n.
\end{equation}
Combining the various results, and defining
\begin{equation}
k(u)=\exp\left (-\int_{u_-}^u\Phi_1'(y)dy\right ),
\end{equation}
gives to leading order in $\epsilon$,
\begin{equation}
\label{WKBf}
\vphi(u)\sim {\mathcal N}k(u)\exp \left (-\frac{\Phi_0(u)}{\epsilon}\right ){\bm \psi}_1(u),
\end{equation}
where $\sum_{n=0}^{\infty} ({\bm \psi}_1)_n=1$ and $ {\mathcal N}$ is a normalization factor, 
\begin{equation}
{\mathcal N}=\left [\int_0^{u_*}k(u)\exp \left (-\frac{\Phi_0(u)}{\epsilon}\right )\right ]^{-1}.
\end{equation}
The latter can be approximated using Laplace's method to give
\begin{equation}
{\mathcal N}\sim \frac{1}{k(u_-)}\sqrt{\frac{|\mu_1'(u_-)|}{2\pi \epsilon}}\exp \left (\frac{\Phi_0(u_-)}{\epsilon}\right )=\sqrt{\frac{|\mu_1'(u_-)|}{2\pi \epsilon}}.
\end{equation}

\subsection{Perturbation analysis of the adjoint eigenfunction}
Following Refs. \cite{Keener11,Newby11,Newby12}, the adjoint eigenfunction ${\bm \xi}_0$ can be approximated using singular perturbation methods. Since $\lambda_0$ is exponentially small in $\epsilon$, equation (\ref{adj}) yields the leading order equation
\begin{equation}
\label{xi0}
\epsilon {\bf V}(u)\frac{\d {\bm \xi}_0}{d u}+{\bf A}^T(u){\bm \xi}_0=0,
\end{equation}
supplemented by the absorbing boundary condition
\begin{equation}
\label{bc1}
({\bm \xi}_0)_n(u_*)=0,\quad n\geq k
\end{equation}
A first attempt at obtaining an approximate solution that also satisfies the boundary conditions is to construct a boundary layer in a neighborhood of the unstable fixed point $u_*$ by performing the change of variables $u=u_*-\epsilon z$ and setting ${\bf Q}(z)={\bf \xi}_0(u_*-\epsilon z)$. Equation (\ref{xi0}) then becomes
\begin{equation}
\label{xi1}
 {\bf V}(u_*)\frac{\d {\bm Q}}{dz}+{\bf A}^T(u_*){\bm Q}=0.
\end{equation}
This inner solution has to be matched with the outer solution ${\bm \xi}_0={\bf 1}$, which means that
\begin{equation}
\lim_{z\rightarrow \infty} {\bf Q}(z)={\bf 1}.
\end{equation}
Recall from equation (\ref{eigM}) that the ($u$--dependent) matrix ${\bf M}={\bf V}^{-1}{\bf A}$ has eigenvalues $\mu_j$. Hence, introducing the similarity transform $\widehat{\bf M}={\bf V}{\bf M}{\bf V}^{-1}$ and taking the transpose shows that $\widehat{\bf M}^T={\bf V}^{-1}{\bf A}^T$ has the same eigenvalues. Denoting the corresponding eigenvectors by ${\bm \zeta}_j$ we introduce the eigenfunction expansion
\begin{equation}
\label{Qz}
{\bf Q}(z)=c_0{\bf 1} +\sum_{j=1}^{\infty}c_j {\bm \zeta}_j(u_*)\e^{\mu_j(u_*)z},
\end{equation}
where
\begin{equation}
\mu_j(u) {\bf V}(u){\bm \zeta}_j(u)+{\bf A}^T(u){\bm \zeta}_j(u)=0.
\end{equation}
In order that the solution remains bounded as $z\rightarrow \infty$ we require that $c_j=0$ if $\mu_j(0)>0$. The boundary conditions (\ref{bc1}) generate a system of  linear equations for the coefficients $c_j$ with codimension $k$. One of the unknowns in determined by matching the outer solution, which suggests that there are $k-1$ positive eigenvalues. The eigenvalues are ordered so that $\mu_j(0)>0$ for $j\geq k-1$. 

There is, however, one problem with the above eigenfunction expansion, namely, that $\mu_1(u_*)=0$ so that the zero eigenvalue is degenerate. Hence, the solution needs to include a secular term involving the generalized eigenvector ${\bm \zeta}_0$,
\begin{equation}
\label{ztop}
{\bf A}^T(u_*){\bm \zeta}_0=-{\bf V}(u_*){\bf 1}= -{\bf v}(u_*).
\end{equation}
The Fredholm alternative ensures that ${\bm \zeta}_0$ exists, since $\rho(u_*)$ is the right null vector of ${\bf A}$ and ${\bf \rho}(u_*)\cdot {\bf v}(u_*)=0$, see equation (\ref{ud}). In component form with $({\bm \zeta}_0)_n=\zeta_n$,
\begin{equation}
 F(u_*)\zeta_{n+1}+n\zeta_{n-1}-(F(u_*)+n)\zeta_n=u_*-wn.
\end{equation}
It is straightforward to show that this has the solution (up to an arbitrary constant that doesn't contribute to the principal eigenvalue)
\begin{equation}
\zeta_n=wn.
\label{zet}
\end{equation}
The solution for ${\bf Q}(z)$ is now
\begin{equation}
\label{Q}
{\bf Q}(z)=c_0{\bf 1} +c_1({\bm \zeta}_0-z{\bf 1})+\sum_{j\geq 2}c_j {\bm \zeta}_j(u_*)\e^{\mu_j(u_*)z}.
\end{equation}
The presence of the secular term means that the solution is unbounded in the limit $z\rightarrow \infty$, which means that the inner solution cannot be matched with the outer solution. One way to remedy this situation is to introduce an alternative scaling in the boundary layer of the form $u=u_*+\epsilon^{1/2}z$,  as detailed in Ref. \cite{Newby12}. Here we simply state the results of the analysis. The full inner solution takes the form
\begin{eqnarray}
{\bm \xi}_0(u)&\sim& \left [1-\hat{c}_1\left (\sqrt{\frac{\pi}{2|\mu_1'(u_*)|}}-\int_{u_*}^{u/\epsilon^{1/2}}\e^{\mu_1'(u_*)y^2}dy\right )\right]{\bf 1}\nonumber \\
&& -\epsilon^{1/2}\hat{c}_1\e^{\mu_1'(u_*)(u-u^*)^2/2\epsilon}{\bm \zeta}_0+\sum_{j\geq 2}\hat{c}_j\e^{\mu_j(u_*)(u-u^*)/\epsilon}{\bm \zeta}_j.
\end{eqnarray}
The remaining coefficients $\hat{c}_1,c_2,\ldots$ are determined by the boundary conditions (\ref{bc1}), which reduce to
\begin{equation}
\label{bcc}
\hat{c}_1 \left (\sqrt{\frac{\pi}{2|\mu_1'(u_*)|}}{\bf 1}+\epsilon^{1/2}{\bm \zeta}_0 \right )_n-\sum_{j\geq 2}\hat{c}_j({\bm \zeta}_j)_n =1
\end{equation}
for $n\geq k$. We thus find that
\begin{equation}
\hat{c}_1 \sim \sqrt{\frac{2|\mu_1'(u_*)|}{\pi}}+{\mathcal O}(\epsilon^{1/2}),\quad \hat{c}_j ={\mathcal O}(\epsilon^{1/2})\, \mbox{for}\, j\geq 2
\end{equation}

\subsection{Principal eigenvalue}
It turns out that we only require the first coefficient $c_1$ in order to evaluate the principal eigenvalue $\lambda_0$ using equation (\ref{l0}). This follows from the observation that ${\bf V}{\bm \eta}_j$ is an eigenfunction of the matrix ${\bf A}^T{\bf V}^{-1}$, which are biorthogonal to the set of eigenvectors ${\bm \psi}_j$ of ${\bf V}^{-1}{\bf A}$. Since the WKB approximation $\vphi$ is proportional to ${\bm \psi}_1$, see equation (\ref{WKBf}), it follows that $\vphi$ is orthogonal to all eigenvectors ${\bm \zeta}_j$, $j\neq 1$. Simplifying the denominator of equation (\ref{l0}) by using the outer solution ${\bm \xi}_0\sim {\bf 1}$, we obtain
\begin{eqnarray}
\label{eq:1}
\lambda_0&\sim& -\frac{{\bm \xi}_0(u_*)^T{\bf V}(u_*)\vphi(u_*)}{\langle \vphi,{\bf 1}\rangle }\nonumber\\
&\sim&c_1{k(u_*)}B(u_*)\sqrt{\frac{|\mu_1'(u_-)|}{2\pi }}\exp\left (-\frac{\Phi_0(u_*)}{\epsilon}\right ),
\end{eqnarray}
with
\begin{eqnarray}
B(u_*)={\bm \zeta}_0^T{\bf V}(u_*){\bf \rho}(u_*)=\sum_{n=0}^{\infty} \zeta_n v_n(u_*)\rho_n(u_*)
\end{eqnarray}
Substituting for $c_1$ and using the relation $\mu_1'(u)=\Phi_0''(u)$,
\begin{eqnarray}
\lambda_0
&\sim&\frac{1}{\pi}{k(u_*)}B(u_*)\sqrt{\Phi_0''(u_-)|\Phi_0(u_*)|}\exp\left (-\frac{\Phi_0(u_*)}{\epsilon}\right ),
\end{eqnarray}
Finally note that $B(u_*)$ can be evaluated using equations (\ref{ss1}) and (\ref{zet}):
\begin{eqnarray}
B(u_*)
&=&w\sum_{n=0}^{\infty}\rho_n(u_*)\left [-u_*n+wn^2\right ]\nonumber \\
&=&w\left [-u_*\langle n\rangle +w\langle n^2\rangle \right ].
\end{eqnarray}
Recall that $\rho_n(u)$ is given by a Poisson density with rate $F(u)$, which implies that $\langle n^2\rangle =\langle n\rangle+\langle n \rangle ^2$ with $\langle n\rangle = F(u_*)$. Therefore, 
\begin{eqnarray}
\label{BB}
B(u_*)&=&{w}\left [wF(u_*)+ F(u_*)(wF(u_*)-u_*) \right ]={w^2F(u_*)}.
\end{eqnarray}

It is instructive to compare the effective potential $\Phi_0$ obtained using the WKB approximation with the potential obtained using the FP equation (\ref{FPC}) based on the QSS approximation. In the one-population case, equations (\ref{DD}) and (\ref{zz}) reduce to
\begin{equation}
\label{DDD}
D=\sum_{n}z_n(u)[v_n(u)-V(u)]\rho_n(u),
\end{equation}
and
\begin{equation}
\sum_n z_n(u)A_{nm}(u)=-[v_m(u)-V(u)].
\end{equation}
Here $v_n$ and ${\bf A}$ are given by equations (\ref{Vcomp}) and (\ref{Acomp}), $\rho_n$ is the Poisson distribution (\ref{poiss}), and $V(u)=-u+wF(u)$. At a fixed point, the equation for $z_n$ reduces to equation (\ref{ztop}), and we find that $z_n=wn$ even away from fixed points. Substituting into equation (\ref{DDD}) shows that
\begin{equation}
D=\langle w^2n[n-F(u)]\rangle = w^2 F(u)\equiv B(u)
\end{equation}
with $B$ given by equation (\ref{BB}). The steady-state solution of the FP equation (\ref{FPC}) takes the form $C(u)\sim \exp^{-\widehat{\Phi}_0(u)/\epsilon}$
with stochastic potential
\begin{equation}
\label{pp1}
\widehat{\Phi}_0(u)=-\int^u\frac{V(y)}{D(y)}dy=-\int^u\frac{-y+wF(y)}{w^2F(y)}dy.
 \end{equation}
Note that $\widehat{\Phi}$ differs from the potential $\Phi_0$, equation (\ref{pot}), obtained using the more accurate WKB method. Equations (\ref{mu1}) and (\ref{pot}) show that the latter has the integral form
\begin{equation}
\label{pp0}
\Phi_0(u)=-\int^u \frac{1}{w}\left [\frac{wF(y)}{y}-1\right ]dy.
\end{equation}
Thus, there will be exponentially large differences between the steady-states for small $\epsilon$. However, it gives the same Gaussian-like behavior close to a fixed point $u_*$, that is,
 \begin{equation}
\left . \frac{\partial \Phi_0}{\partial u}\right |_{u=u_*}=\left . \frac{\partial \widehat{\Phi}_0}{\partial u}\right |_{u=u_*}=0,\quad \left . \frac{\partial^2 \Phi_0}{\partial u^2}\right |_{u=u_*}=\left . \frac{\partial^2 \widehat{\Phi}_0}{\partial u^2}\right |_{u=u_*}=\left . \frac{1-wF'(u)}{wu}\right |_{u=u_*}
\end{equation}

\subsection{Results}

\begin{figure}[h!]
  \centering
  \includegraphics[width=10cm]{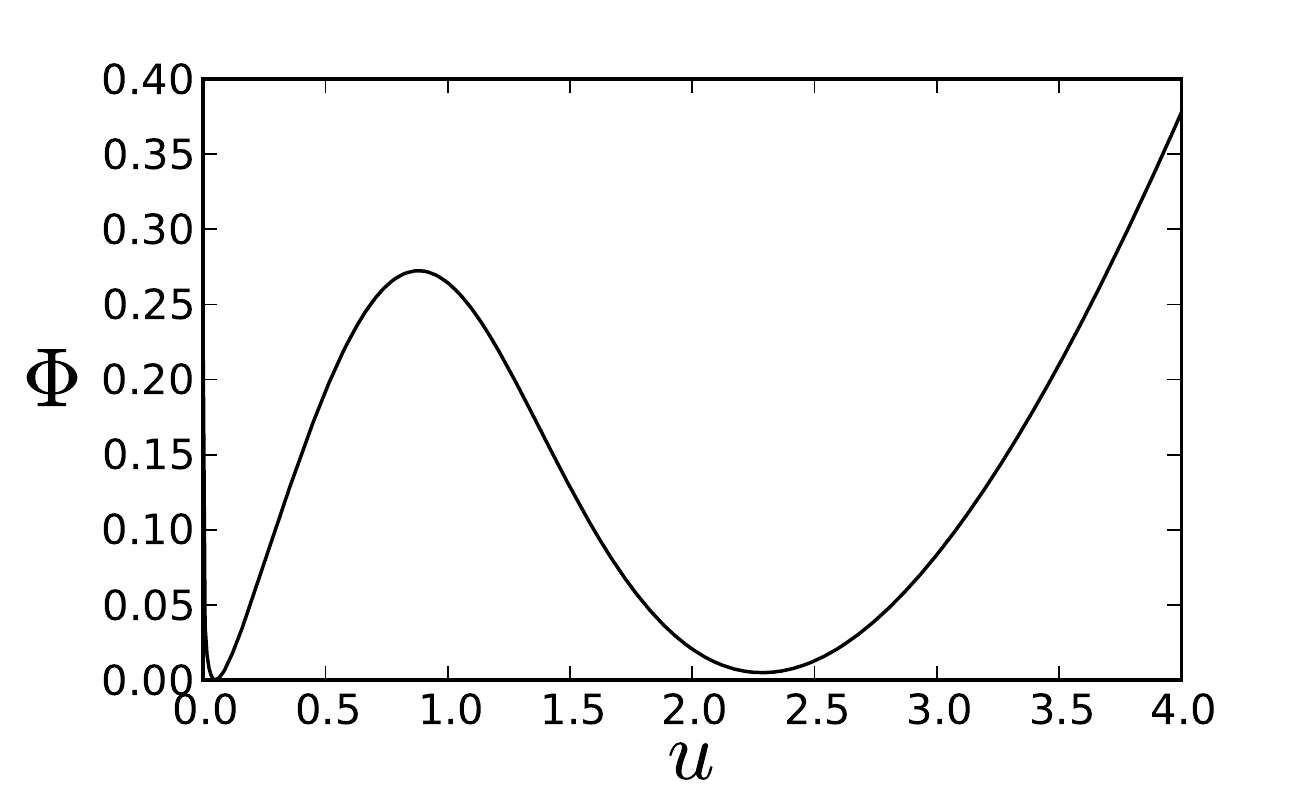}
  \caption{Comparison of the double-well potentials $\Phi_0(u)$ and $\widehat{\Phi}_0(u)$ obtained using the quasistationary approximation and the QSS diffusion approximation, respectively. Parameter values are  chosen so that deterministic network is bistable: $F_{0}=2$, $\gamma=4$, $\kappa=1$, and $w=1.15$.}
  \label{fig:potential}
\end{figure}

\begin{figure}[b!]
  \centering
  \includegraphics[width=12cm]{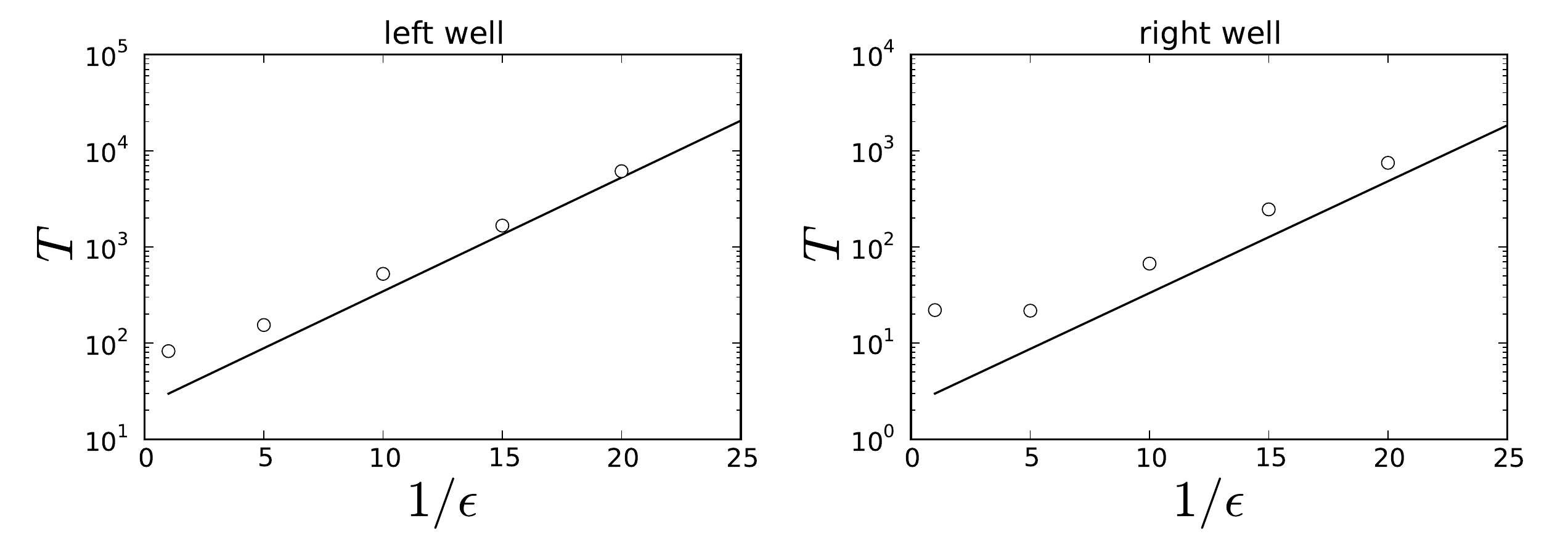}
  \caption{Mean exit time from the left and right well calculated using the quasistationary approximation (solid line) and the QSS diffusion approximation (dashed line).  The open circles represent data points obtained by numerically solving the corresponding jump velocity Markov process using the Gillespie algorithm. Parameter values are the same as in Fig.~\ref{fig:potential}.}
  \label{fig:1Dmfpt}
\end{figure}

In Fig.~\ref{fig:potential}, we plot the potential function $\Phi_0$ of equation (\ref{pp0}), which is obtained using the quasistationary approximation in a parameter regime for which the underlying deterministic network is bistable. We also plot the corresponding potential function $\widehat{\Phi}_0$ of equation (\ref{pp1}), under the QSS diffusion approximation. The differences between the two lead to exponentially large differences in estimates for the mean exit times when $\epsilon$ is small. The mean exit time from the left and right well is shown in Fig.~\ref{fig:1Dmfpt}. Solid curves show the analytical approximation $T\sim 1/\lambda_{0}$, where $\lambda_{0}$ is given by \eqref{eq:1}, as a function of $1/\epsilon$.  For comparison, the mean exit time computed from averaged Monte-Carlo simulations of the full stochastic system are shown as symbols. From \eqref{eq:1}, we expect the log of the mean exit time to be an asymptotically-linear function of $1/\epsilon$, and this is confirmed by Monte-Carlo simulations.  The slope is determined by the depth of the potential well, and the vertical shift is determined by the prefactor. Also shown is the corresponding MFPT calculated using the QSS diffusion approximation (dashed curves), which is typically several orders of magnitude out, and validates the relative accuracy of the quasistationary approximation.

\section{Metastable states in a two population model}

\begin{figure}[htbp]
\begin{center}
\includegraphics[width=7cm]{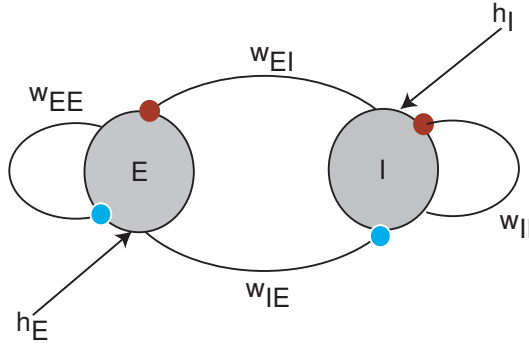}
\caption{\small Two--population E-I network with both intrasynaptic connections $w_{EE},w_{II}$ and intersynaptic connections $w_{IE},w_{EI}$. There could also be external inputs $h_E,h_I$, which can be incorporated into the rate functions of the two-populations by shifting the firing threshold $\kappa$.}
\label{gain}
\end{center}
\end{figure}

The same basic analytical steps as \S 3 can also be used to study multipopulation models ($M>1$). One now has $M$ piecewise-deterministic variables $U_{\alpha}$ and $M$ discrete stochastic variables $N_{\alpha}$ evolving according to the CK equation (\ref{CK}). The mean first passage time is again determined by the principal eigenvalue $\lambda_0$ of the corresponding linear operator. As in the one-population model, $\lambda_0$ can be expressed in terms of inner products involving a quasi stationary density $\vphi$, obtained using a multidimensional WKB method, and the principal eigenvector $\xi_0$ of the adjoint linear operator, calculated using singular perturbation theory. One of the major differences between the one-population model and multi-dimensional versions is that the latter exhibit much richer dynamics in the mean-field limit, including oscillatory solutions. For example, consider a two-population model ($M=2$) consisting of an excitatory population interacting with an inhibitory population as shown in Fig. \ref{gain}. This is one of the simplest deterministic networks known to generate limit cycle oscillations at the population level  \cite{Borisyuk92}, and figures as a basic module in many population models. For example, studies of stimulus--induced oscillations and synchrony in primary visual cortex often take the basic oscillatory unit to be an E-I network operating in a limit cycle regime \cite{Schuster90,Grannan93}. Here the E-I network represents a cortical column, which can synchronize with other cortical columns either via long-range synaptic coupling or via a common external drive. In this paper, we will focus on parameter regimes where the two-population model exhibits bistability.

\begin{figure}[htbp]
\begin{center}
\includegraphics[width=10cm]{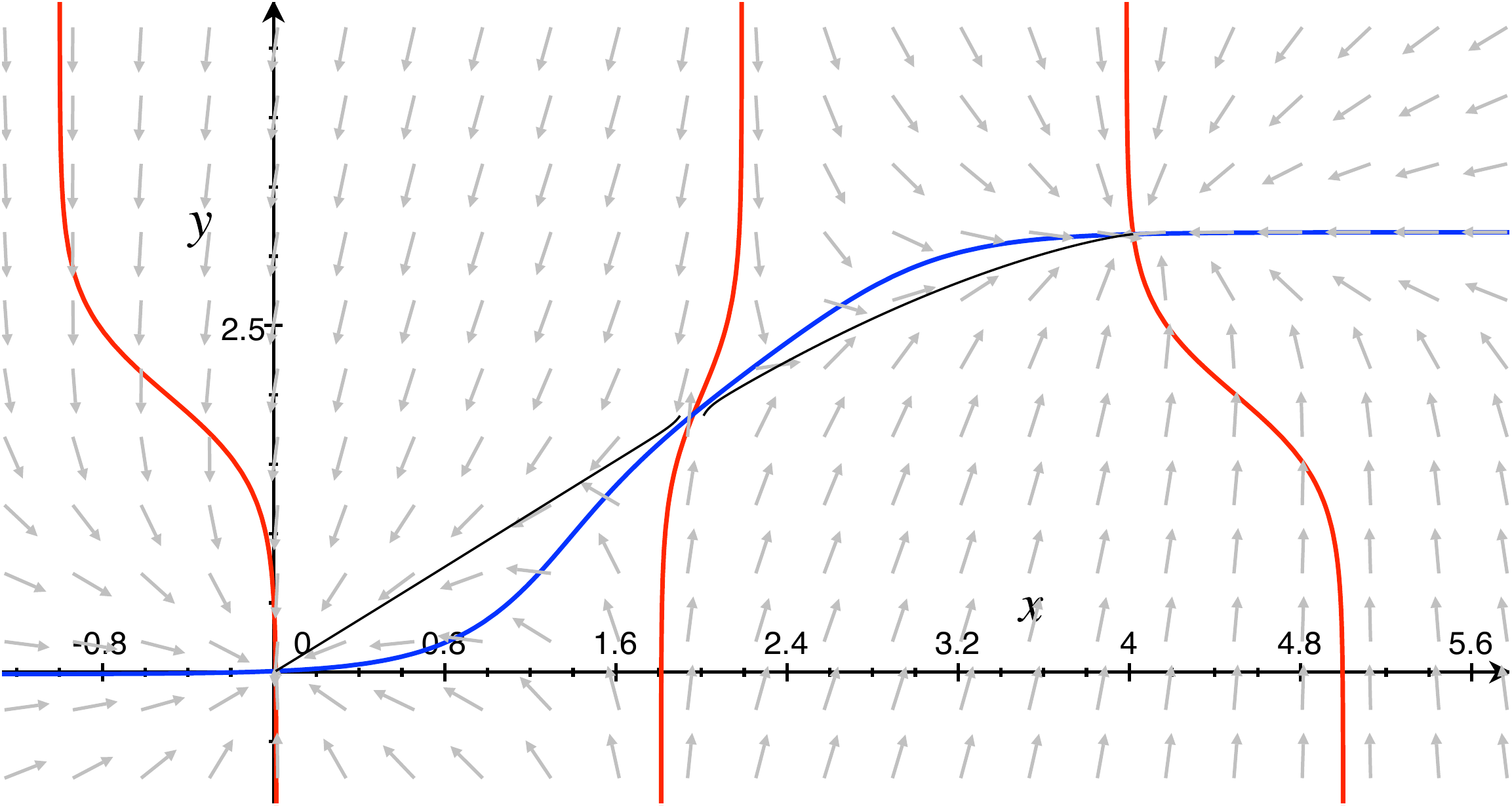}
\caption{\small Deterministic limit of the two population model, showing bistability.  Red curves show the $x$-nullclines, and blue curve show the $y$-nullcline. The red nullcline through the saddle is its stable manifold and acts as the separatrix $\Gamma$ between the two stable fixed points. Two deterministic trajectories are shown (black curves), starting from either side of the unstable saddle and ending at a stable fixed point.  Parameter values are $F_{0}=1$, $\gamma=3$, $\kappa=2$, $w_{EE}=5$, $w_{EI} = 1$, $w_{IE} = 9$, and $w_{II} = 6$. }
\label{fig:bistable2}
\end{center}
\end{figure}

Let $u_1=x$ and $u_2=y$ denote the synaptic variables of the excitatory and inhibitory networks, respectively, and denote the corresponding spiking variables by $n_x$ and $n_y$. The CK equation (\ref{CK}) can be written out fully as
\begin{eqnarray}
\label{CK2}
\frac{\partial p}{\partial t}&=&-\frac{\partial (v p)}{\partial x}-\frac{\partial (\widetilde{v}p)}{\partial y}\\
&&+\frac{1}{\epsilon}\left [ \omega_-(n_x+1)p(x,y,n_x+1,n_y,t) +\omega_-(n_y+1)p(x,y,n_x,n_y+1,t) \right ]\nonumber \\
&&+\frac{1}{\epsilon}\left [ \omega_+(x)p(x,y,n_x-1,n_y,t) +\omega_+(y)p(x,y,n_x,n_y-1,t) \right ]\nonumber \\
&&-\frac{1}{\epsilon}\left [ \omega_-(n_x)+\omega_-(n_y)+\omega_+(x)+\omega_+(y) \right ]p(x,y,n_x,n_y,t),\nonumber 
\end{eqnarray}
where
\begin{eqnarray}
\label{v2}
v(x,n_x,n_y)&=&-x+\left [w_{EE}n_x-w_{EI}n_y\right ],\\
 \widetilde{v}(y,n_x,n_y)&=&-y+\left [w_{IE}n_x-w_{II}n_y\right ],
 \label{v2til}
\end{eqnarray}
and
\begin{equation}
\omega_+(x)=F(x),\quad \omega_-(n)=n.
\end{equation}
Thus the synaptic coupling between populations occurs via the drift terms $v,\widetilde{v}$. As in the case of the one--population model, we expect the finite-time behavior to be characterized by small Gaussian fluctuations about the stable steady-state of the corresponding pure birth-death process. We now show that in the limit $N\rightarrow \infty$ and $\Delta \tau\rightarrow 0$ with $N\Delta t=1$ and $\epsilon$ fixed, the steady-state distribution reduces to a multivariate Poisson process. First, introduce the generating function (for fixed $(x,y)$)
\begin{equation}
G(r,s)=\sum_{n_y=0}^{\infty}\sum_{n_y=0}^{\infty} r^{n_x}s^{n_y}p(x,y,n_x,n_y).
\end{equation}
Setting all derivatives in equation (\ref{CK2}) to zero, multiplying both sides by $r^{n_x}$ and $r^{n_y}$ and summing over $n_x,n_y$ gives the quasilinear equation
\begin{equation}
0=(1-r)\frac{\partial G}{\partial r}+(1-s)\frac{\partial G}{\partial s}+[(r-1)\omega_+(x)+(s-1)\omega_+(y)]G.
\end{equation}
This can be solved using the method of characteristics to give
\begin{equation}
G(r,s)=\exp\left ([r-1]\omega_+(x)+[s-1]\omega_+(y)\right ],
\end{equation}
which is the generating function for the steady--state Poisson distribution
\begin{equation}
\label{Poi}
\rho(x,y,n_x,n_y)=\frac{\omega_+(x)^{n_x}\e^{-\omega_+(x)}}{n_x!}\cdot \frac{\omega_+(y)^{n_y}\e^{-\omega_+(y)}}{n_y!}.
\end{equation}
Since $\langle n_x\rangle =\omega_+(x) ,\langle n_y\rangle =\omega_-(y)$, it immediately follows that in the limit $\epsilon \rightarrow 0$, we obtain the standard voltage-based mean-field equations for an E-I system:
\begin{eqnarray}
\label{rateEI}
\label{xdot}
\frac{dx}{dt}&=&-x+w_{EE}F(x)-w_{EI}F(y) , \\
\frac{dy}{dt}&=&-y+w_{IE}F(x)-w_{II}F(y)
\label{ydot}
\end{eqnarray}
It is well known that the dynamical system (\ref{rateEI}) exhibits multistability and limit cycle oscillations \cite{Borisyuk92}. We will assume that it is operating in a parameter regime for which there is bistability as illustrated in Fig. \ref{fig:bistable2}. For example, if $w_{EE}-w_{EI}=w_{IE}-w_{II}=w$ then $x=y$ is an invariant manifold on which
$x$ evolves according to the one-population equation (\ref{rate}). Varying the threshold $\kappa$ then leads to a pitchfork bifurcation and the emergence of bistability.

In order to analyze the effects of fluctuations for $0<\epsilon \ll 1$, we rewrite equation (\ref{CK2}) in a more compact form by introducing some tensor notation. First, we introduce the probability 1-tensor $\p(x,y,t)$ with components 
\begin{equation}
p_{n_x,n_y}(x,y,t)=p(x,y,n_x,n_y,t)
\end{equation}
the diagonal drift 2-tensors ${\bf V}(x),\widetilde{\bf V}(y)$ with diagonal components 
\begin{equation}
\label{VV2}
V_{n_x,n_y;n_x,n_y}(x)=v(x,n_x,n_y),\quad \widetilde{V}_{n_x,n_y;n_x,n_y}(y)=\widetilde{v}(y,n_x,n_y),
\end{equation}
and the transition 2-tensor ${\bf A}(x,y)$ with non-zero components
\begin{equation}
A_{n_x,n_y;n_x-1,n_y}(x,y)=F(x),\quad A_{n_x,n_y;n_x,n_y-1}(x,y)=F(y),
\end{equation}
\begin{equation}
A_{n_x,n_y;n_x+1,n_y}(x,y)=n_x+1,\quad A_{n_x,n_y;n_x,n_y+1}(x,y)=n_y+1,
\end{equation}
and
\begin{equation}
A_{n_x,n_y;n_x,n_y}(x,y)=-[F(x)+F(y)+n_x+n_y].
\end{equation}
Second, we rewrite the CK equation as
\begin{equation}
\label{CK2v}
\frac{\partial \p}{\partial t}=-\frac{\partial}{\partial x}({\bf V}\circ \p)-\frac{\partial}{\partial y}(\widetilde{\bf V}\circ \p)+\frac{1}{\epsilon}{\bf A}\circ {\bf p},
\end{equation}
where 
\begin{equation}
[{\bf A}\circ {\p}]_{n_x,n_y}=\sum_{m_x,m_y}A_{n_x,n_y;m_x,m_y}p_{m_x,m_y},
\end{equation}
etc. The tensor ${\mathbf A}$ satisfies the null constraints (cf. equation (\ref{null})
\begin{equation}
{\bf 1}^T\circ {\bf A}=0,\quad {\bf A}\circ {\bm \rho}=0,
\end{equation}
with ${\bf 1}_{n_x,n_y}=1$ for $n_x,n_y$ and $\rho_{n_x,n_y}(x,y)=\rho(x,y,n_x,n_y)$ with the density $\rho$ given by the Poisson distribution (\ref{Poi}). In tensor notation, the mean-field equations (\ref{rateEI}) can be written as
\begin{equation}
\frac{dx}{dt}={\bf V}\circ {\bm \rho},\quad \frac{dy}{dt}=\widetilde{\bf V}\circ {\bm \rho}.
\end{equation}
Finally, note that as in the one-population model, we can restict the domain of the stochastic dynamics in the $(x,y)$--plane. In order to show this, multiply both sides of equations (\ref{v2}) and (\ref{v2til}) by $w_{II}$ and $w_{EI}$ respectively, and add the resulting equations. Setting 
\begin{equation}
\label{hatx}
\hat{x}=[w_{II}x-w_{EI}y]/\det[{\bf w}]
\end{equation}
with $\det[{\bf w}]=w_{EE}w_{II}-w_{EI}w_{IE}$, we have the transformed drift term
\begin{equation}
\hat{v}(\hat{x},n_x)= -\hat{x}+{n_x}.
\end{equation}
Similarly, multiplying both sides of equations (\ref{v2}) and (\ref{v2til}) by $w_{IE}$ and $w_{EE}$ respectively, and adding the resulting equations yields
\begin{equation}
\hat{\widetilde{v}}(\hat{y},n_y)= -\hat{y}+{n_y}.
\end{equation}
with 
\begin{equation}
\label{haty}
\hat{y}=[w_{IE}x-w_{EE}y]/\det[{\bf w}].
\end{equation}
It follows that the dynamics can be restricted to the domain $\hat{x}>0,\hat{y}>0$.

\subsection{Quasi-stationary approximation}

In order to investigate rare transitions between the metastable states shown in Fig. \ref{fig:bistable2}, we introduce an absorbing boundary along the separatrix $\Gamma$ separating the two states:
\begin{equation}
\p(x,y,t)=0,\quad (x,y)\in \Gamma
\end{equation}
for all components $(n_x,n_y)$ for which 
\begin{equation}
\label{ls}
(v(x,n_x,n_y),\widetilde{v}(y,n_x,n_y))\cdot \widehat{\bf s}<0,
\end{equation}
where $\hat{\bf s}$ is the unit normal of $\Gamma$ pointing into the domain ${\mathcal D}$ of the initial metastable state, which we take to be $(x_-,y_-)$. Following identical arguments to the one-population model, we can expand the probability density as
\begin{equation}
\p(x,y,t)=\sum_{j=0}^{\infty}c_j{\bm \phi}_j(x,y)\e^{-\lambda_j t}
\end{equation}
with $(\lambda_j,{\bm \phi}_j)$ determined from the eigenvalue equation
\begin{equation}
\label{Lhat2}
\widehat{L}{\bm \phi}_j\equiv \frac{\partial }{\partial x}({\bf V}\circ {\bm \phi}_j)+\frac{\partial }{\partial y}(\widetilde{\bf V}\circ{\bm \phi}_j)-\frac{1}{\epsilon}{\bf A}\circ {\bm \phi}_j=\lambda_j{\bm \phi}_j,
\end{equation}
together with the boundary conditions
\begin{equation}
({\bm \phi}_j)(x,y,n_x,n_y)=0,\, \mbox{for all}\, (x,y)\in \Gamma,\, \mbox{and}\, (n_x,n_y)
\end{equation}
for which equation (\ref{ls}) is satisfied.
The principal eigenvalue $\lambda_0$ again determines the first passage time density according to $f(t)\sim \lambda_0\e^{-\lambda_0 t}$. Moreover, $\lambda_0$ can be approximated using a spectral projection method that makes use of the adjoint eigenvalue equation
\begin{equation}
\widehat{L}^*{\bm \xi_j}\equiv -{\bf V}\circ \frac{\partial }{\partial x}({\bm \phi}_j)-\widetilde{\bf V}\circ \frac{\partial }{\partial y}({\bm \phi}_j)-\frac{1}{\epsilon}{\bf A}\circ{\bm \xi}_j=\lambda_j{\bm \xi}_j,
\end{equation}
with $\langle {\bm \phi}_i,{\bm \xi}_j\rangle =\delta_{i,j}$ and the inner product defined for 1-tensors according to
\begin{eqnarray}
\langle {\bm f},{\bm g}\rangle&=&\int_{\mathcal D}{\bm f}(x,y)^T\circ {\bm g}(x,y)dxdy,\nonumber \\
&=&\sum_{n_x,n_y}\int_{\mathcal D}f(x,y,n_x,n_y)g(x,y,n_x,n_y)dxdy.
\end{eqnarray}
Now suppose that we replace ${\bm \phi}_0$ by the quasi-stationary density $\vphi$, for which $\widehat{L}\vphi=0$ without satisfying the absorbing boundary conditions. Application of the divergence theorem shows that
\begin{equation}
\langle \vphi,\widehat{L}^*{\bm \xi}_0\rangle = \langle \widehat{L} \vphi,{\bm \xi}_0\rangle+\int_{\Gamma}\left ({\bm \xi}_0^T\circ{\bf V}\circ \vphi , \,{\bm \xi}_0^T\circ \widetilde{\bf V}\circ \vphi \right )\cdot \widehat{\bf n}ds.
\end{equation}
It follows that
\begin{equation}
\label{lam02}
\lambda_0=\frac{\left ({\bm \xi}_0^T\circ{\bf V}\circ \vphi ,\, {\bm \xi}_0^T\circ \widetilde{\bf V}\circ \vphi \right )\cdot \widehat{\bf n}ds}{\int_{\mathcal D}{\bm \xi}_0^T\circ \vphi dA}.
\end{equation}

\subsection{WKB method and the quasi-stationary density}
Following along similar lines to the one-population model, we approximate the quasi-stationary density $\vphi(x,y)$ of the CK equation (\ref{CK2v}) using the WKB method. That is, we seek an approximate solution of $\widehat{L}{\bm \phi}_{\epsilon}=0$ of the WKB form
\begin{equation}
\label{WKB2}
\vphi(u)=\left [{\bf R}_0(x,y)+\epsilon {\bf R}_1(x,y)\right ]\exp\left (-\frac{\Phi_0(x,y)+\epsilon \Phi_1(x,y)}{\epsilon}\right ),
\end{equation}
Here ${\bf R}_0$ and ${\bf R}_1$ are 1-tensors and $\Phi_0,\Phi_1$ are scalars.
Substituting into equation (\ref{CK2v}) and collecting leading-order terms in $\epsilon$ gives
\begin{equation}
\label{wow}
[{\bf A}+{\mathcal P}_x{\bf V}+{\mathcal P}_y\widetilde{\bf V}]\circ {\bf R}_0=0,
\end{equation}
where
\begin{equation}
\label{Psig}
{\mathcal P}_x=\frac{\partial \Phi_0}{\partial x},\quad {\mathcal P}_y=\frac{\partial \Phi_0}{\partial y}.
\end{equation}
Given the explicit form of the diagonal tensors ${\bf V},\widetilde{\bf V}$, see equation (\ref{VV2}), ${\mathcal P}_1{\bf V}+{\mathcal P}_2\widetilde{\bf V}$ for $(x,y)\in {\mathcal D}$ has at least two components of opposite sign. This is a necessary condition for the existence of a non-trivial positive solution for ${\mathbf R}_0$ in the domain ${\mathcal D}$ according to Theorem 3.1 of \cite{Keener11} . 

We now make the ansatz
\begin{equation}
[{\bf R}_0]_{n_x,n_y}=\frac{\Lambda_x^{n_x}}{n_x!}\cdot \frac{\Lambda_y^{n_y}}{n_y!}.
\end{equation}
Substituting into equation (\ref{wow}) and using the explicit expressions for ${\mathbf A}$, ${\bf V}$ and $\widetilde{\bf V}$, we find that
\begin{eqnarray}
\label{boo}
&&\left [\frac{F(x)}{\Lambda_x}-1\right ]n_x+\left [\frac{F(y)}{\Lambda_y}-1\right ]n_y+\Lambda_x+\Lambda_y-F(x)-F(y)\nonumber \\
&&\quad =-{\mathcal P}_x[-x+w_{EE}n_x-w_{EI}n_y]-{\mathcal P}_y[-y+w_{IE}n_x-w_{II}n_y].
\end{eqnarray}
The variables ${\mathcal P}_x$ and ${\mathcal P}_y$ can be determined by cancelling terms in $n_x$ and $n_y$. This yields the pair of simultaneous equations
 \begin{eqnarray}
 \frac{F(x)}{\Lambda_x}-1&=&-[w_{EE}{\mathcal P}_x+w_{IE}{\mathcal P}_y],\quad
 \frac{F(y)}{\Lambda_y}-1=w_{EI}{\mathcal P}_x+w_{II}{\mathcal P}_y.
 \label{bot}
 \end{eqnarray}
 Substituting back into equation (\ref{boo}) gives
 \begin{equation}
 \label{goo}
x{\mathcal P}_x+y{\mathcal P}_y=\Lambda_x+\Lambda_y-F(x)-F(y).
 \end{equation}
Solving for $\Lambda_x,\Lambda_y$ in terms of ${\mathcal P}_x$ and ${\mathcal P}_y$, equation (\ref{goo}) can be rewritten as 
 \begin{equation}
\label{maH}
{\mathcal H}(x,y,{\mathcal P}_x,{\mathcal P}_y)\equiv -x{\mathcal P}_x-y{\mathcal P}_y -F(x)-F(y)+\Lambda_x(x,{\mathcal P}_x,{\mathcal P}_y)+\Lambda_y(y,{\mathcal P}_x,{\mathcal P}_y)=0,
\end{equation}
where
\begin{eqnarray}
\Lambda_x=\frac{F(x)}{1-w_{EE}{\mathcal P}_x-w_{IE}{\mathcal P}_y},\quad
\Lambda_y=\frac{F(y)}{1+w_{EI}{\mathcal P}_x+w_{II}{\mathcal P}_y}
 \end{eqnarray}
Mathematically speaking, equation (\ref{maH}) is identical in form to a stationary Hamilton Jacobi equation for a classical particle moving in the domain ${\mathcal D}$, with ${\mathcal H}$ identified as the Hamiltonian. A trajectory of the particle is given by the solution of Hamilton's equations
\begin{eqnarray}
\label{eq:3}
&& \frac{dx}{dt}=\frac{\partial {\mathcal H}}{\partial {\mathcal P}_x},\quad \frac{dy}{dt}=\frac{\partial {\mathcal H}}{\partial {\mathcal P}_y},\nonumber \\
&& \frac{d\mathcal P_x}{dt}=-\frac{\partial {\mathcal H}}{\partial x},\quad \frac{d\mathcal P_y}{dt}=-\frac{\partial {\mathcal H}}{\partial y}
\end{eqnarray}
Here $t$ is treated as a parameterization of trajectories rather than as a real time variable. Given a solution curve $(x(t),y(t))$, known as a ray, the potential $\Phi_0$ can be determined along the ray by solving the equation
\begin{equation}
\frac{d\Phi_0}{dt}\equiv \frac{\partial \Phi_0}{\partial x}\frac{dx}{dt}+ \frac{\partial \Phi_0}{\partial y}\frac{dy}{dt}={\mathcal P}_x\frac{dx}{dt}+ {\mathcal P}_y\frac{dy}{dt}.
\end{equation}
Thus, $\Phi_0$ can be identified as the action along a zero energy trajectory. One can then numerically solve for $\Phi_0$ by considering Cauchy data in a neighborhood of the stable fixed point $(x_-,y_-)$ \cite{Newby12}. We find that Hamilton's equations take the explicit form
\begin{eqnarray}
\label{eq:4}
\frac{dx}{dt}&=&-x+w_{EE}F(x)-w_{EI}F(y)+\frac{w_{EE}{\mathcal P}_x+w_{IE}{\mathcal P}_y}{[1-w_{EE}{\mathcal P}_x-w_{IE}{\mathcal P}_y]^2}w_{EE}F(x)\nonumber 
\\ && -\frac{w_{EI}{\mathcal P}_x+w_{II}{\mathcal P}_y}{[w_{EI}{\mathcal P}_x+w_{II}{\mathcal P}_y+1]^2}w_{EI}F(y)
\label{HHx}
\end{eqnarray}
\begin{eqnarray}
\frac{dy}{dt}&=&-y+w_{IE}F(x)-w_{II}F(y)+\frac{w_{EE}{\mathcal P}_x+w_{IE}{\mathcal P}_y}{[1-w_{EE}{\mathcal P}_x-w_{IE}{\mathcal P}_y]^2}w_{IE}F(x)\nonumber 
\\ && -\frac{w_{EI}{\mathcal P}_x+w_{II}{\mathcal P}_y}{[w_{EI}{\mathcal P}_x+w_{II}{\mathcal P}_y+1]^2}w_{II}F(y)
\label{HHy}
\end{eqnarray}
\begin{eqnarray}
\frac{d\mathcal P_x}{dt}={\mathcal P}_x -\frac{w_{EE}{\mathcal P}_x+w_{IE}{\mathcal P}_y}{1-w_{EE}{\mathcal P}_x-w_{IE}{\mathcal P}_y}F'(x)
\end{eqnarray}
\begin{eqnarray}
\frac{d\mathcal P_y}{dt}={\mathcal P}_y +\frac{w_{EI}{\mathcal P}_x+w_{II}{\mathcal P}_y}{w_{EE}{\mathcal P}_x+w_{IE}{\mathcal P}_y+1}F'(y)
\end{eqnarray}
Note that we recover the mean-field equations along the manifold ${\mathcal P}_x={\mathcal P}_y=0$ with $\Lambda_x=F(x),\Lambda_y=F(y)$.

It remains to specify Cauchy data for the effective Hamiltonian system. At the stable fixed point, the value of each variable is known with ${\mathcal P}_x={\mathcal P}_y=0$ and $(x,y)=(x_-,y_-)$. However, data at a single point is not sufficient to generate a family of rays. Therefore, as is well known in the application of WKB methods \cite{Matkowsky90,Stein97,Schuss10}, it is necessary to specify data on a small ellipse surrounding the fixed point. Thus, Taylor expanding $\Phi_0$ around the fixed point yields, to leading order, the quadratic form
\begin{equation}
\Phi_0(x,y)\approx \frac{1}{2} {\bf z}^T{\bf Z}{\bf z},\quad {\bf z}=\left ( \begin{array}{c} x-x_-\\ y-y_- \end{array}\right ).
\end{equation}
where ${\bf Z}$ is the Hessian matrix
\begin{equation}
{\bf Z}=\left (\begin{array}{cc} \frac{\partial^2 \Phi_0}{\partial x^2} & \frac{\partial^2 \Phi_0}{\partial x\partial y} \\
 \frac{\partial^2 \Phi_0}{\partial y \partial x} & \frac{\partial^2 \Phi_0}{\partial y^2}\end{array} \right ),
 \end{equation}
and we have chosen $\Phi_0(x_-,y_-)=0$. Cauchy data are specified on the $\theta$-parameterized ellipse
\begin{equation}
\frac{1}{2} {\bf z}^T(\theta){\bf Z}{\bf z}(\theta)=\chi,
\end{equation}
for a suitably chosen parameter $\chi$ that is small enough to generate accurate numerical results, but large enough so that the ellipse can generate trajectories that cover the whole domain ${\mathcal D}$. On the elliptical Cauchy curve, the initial values of ${\mathcal P}_x$ and ${\mathcal P}_y$ are
\begin{equation}
\left ( \begin{array}{c} {\mathcal P}_{x,0}(\theta) \\ {\mathcal P}_{y,0}(\theta) \end{array}\right )={\bf Z}\left ( \begin{array}{c} x_0(\theta)-x_-\\ y_0(\theta)-y_- \end{array}\right ).
\end{equation}
It can be shown that the Hessian matrix satisfies the alebraic Riccati equation \cite{Stein97}
\begin{equation}
{\bf Z}{\bf B}{\bf Z}+{\bf Z}{\bf C}+{\bf C}^T{\bf Z}=0,
\end{equation}
where
\begin{equation}
{\bf B}=\left (\begin{array}{cc} \frac{\partial^2{\mathcal H}}{\partial {\mathcal P}_x^2} & \frac{\partial^2{\mathcal H}}{\partial {\mathcal P}_x\partial {\mathcal P}_y} \\
 \frac{\partial^2 {\mathcal H}}{\partial {\mathcal P}_y \partial {\mathcal P}_x} & \frac{\partial^2 {\mathcal H}}{\partial {\mathcal P}_y^2}\end{array} \right ), 
 \quad {\bf C}=\left (\begin{array}{cc} \frac{\partial^2{\mathcal H}}{\partial {\mathcal P}_x \partial x} & \frac{\partial^2{\mathcal H}}{\partial {\mathcal P}_x\partial y} \\
 \frac{\partial^2 {\mathcal H}}{\partial {\mathcal P}_y \partial x} & \frac{\partial^2 {\mathcal H}}{\partial {\mathcal P}_y \partial y}\end{array} \right )
 \end{equation}
 are evaluated at ${\mathcal P}_x={\mathcal P}_y=0$ and $(x,y)=(x_-,y_-)$. In order to numerically solve the Ricatti equation it is convenient to transform into a linear problem by making the substitution ${\bf Q}={\bf Z}^{-1}$:
 \begin{equation}
 {\bf B}+{\bf CQ}+{\bf Q}{\bf C}^T=0.
 \end{equation}

Proceeding to the next order in the WKB solution of equation (\ref{CK2v}), we find that
\begin{equation}
 [{\bf A}+{\mathcal P}_x{\bf V}+{\mathcal P}_y\widetilde{\bf V}]\circ {\bf R}_1=\frac{\partial [{\bf V}\circ {\bf R}_0]}{\partial x}+\frac{\partial [\widetilde{\bf V}\circ {\bf R}_0]}{\partial y}-\left (\frac{\partial \Phi_1}{\partial x}{\bf V}+\frac{\partial \Phi_1}{\partial y}\widetilde{\bf V}\right )\circ{\bf R}_0.
\end{equation} 
Since the 2-tensor ${\bf M}\equiv {\bf A}+{\mathcal P}_x{\bf V}+{\mathcal P}_y\widetilde{\bf V}$ has the unique right null 1-tensor ${\bf R}_0$, it's left null-space is also one-dimensional spanned by ${\bm \eta}$, say. The Fredholm alternative theorem then requires that
\begin{equation}
\label{ella}
{\bm \eta}^T\circ \left [\frac{\partial [{\bf V}\circ {\bf R}_0]}{\partial x}+\frac{\partial [\widetilde{\bf V}\circ {\bf R}_0]}{\partial y}-\left (\frac{\partial \Phi_1}{\partial x}{\bf V}+\frac{\partial \Phi_1}{\partial y}\widetilde{\bf V}\right )\circ {\bf R}_0\right ]=0.
\end{equation}
Using the fact that $({\bm \eta}^T\circ {\bf V}\circ {\bf R}_0)dy/dt=({\bm \eta}^T\circ \widetilde{\bf V}\circ {\bf R}_0)dx/dt$ along trajectories of the Hamiltonian system, we can rewrite the above equation as (cf. equation (\ref{bat}))
\begin{equation}
\frac{d\Phi_1}{dt} \equiv \frac{\partial \Phi_1}{\partial x}\frac{dx}{dt}+ \frac{\partial \Phi_1}{\partial y}\frac{dy}{dt}=\frac{dx/dt}{{\bm \eta}^T\circ{\bf V}\circ{\bf R}_0}{\bm \eta}^T\circ \left (\frac{\partial [{\bf V}\circ{\bf R}_0]}{\partial x}+\frac{\partial [\widetilde{\bf V}\circ{\bf R}_0]}{\partial y}\right ).
\end{equation}
As shown in appendix A of \cite{Newby12}, an equation of this form can be numerically integrated along the trajectories of the underlying Hamiltonian system. 

 However, ${\bm \eta}$ may be solved explicitly by substituting the ansatz
\begin{equation}
\eta_{n_x,n_y}=\Gamma_x^{n_x}\cdot \Gamma_y^{n_y}
\end{equation}
into the equation ${\bm \eta}^T\circ{\bf M}=0$, and using the explicit expressions for ${\bf A}, {\bf V}, \widetilde{\bf V}$. One finds that 
\begin{eqnarray}
\label{boo2}
&&{F(x)}[{\Gamma_x}-1]+{F(y)}[{\Gamma_y}-1]+n_x\left [\frac{1}{\Gamma_x}-1\right ]+n_y\left [\frac{1}{\Gamma_y}-1\right ]\nonumber \\
&&\quad =-{\mathcal P}_x[-x+w_{EE}n_x-w_{EI}n_y]-{\mathcal P}_y[-y+w_{IE}n_x-w_{II}n_y].
\end{eqnarray}
Cancelling the terms in $n_x$ and $n_y$ yields
\begin{eqnarray}
\frac{1}{\Gamma_x}-1&=&-[w_{EE}{\mathcal P}_x+w_{IE}{\mathcal P}_y],\\
 \frac{1}{\Gamma_y}-1&=&w_{EI}{\mathcal P}_x+w_{II}{\mathcal P}_y
\end{eqnarray}
Comparison with equation (\ref{bot}) shows that
\begin{equation}
\Gamma_x=\frac{\Lambda_{x}}{F(x)}, \quad \Gamma_y=\frac{\Lambda_{y}}{F(y)},
\end{equation}
so that
\begin{equation}
\eta_{n_x,n_y}=\left (\frac{\Lambda_x}{F(x)}\right )^{n_x}\cdot \left (\frac{\Lambda_y}{F(y)}\right )^{n_y}.
\end{equation}

In summary,
the quasi-stationary approximation takes the form
\begin{equation}
\label{quiz}
\vphi(x,y)\sim {\mathcal N} \e^{-\Phi_1(x,y)}\e^{-\Phi_0(x,y)/\epsilon}{\bf R}_0.
\end{equation}
The normalization factor ${\mathcal N}$ can be approximated using Laplace's method to give
\begin{equation}
\label{norm2}
{\mathcal N}=\left [\int_{\mathcal D} \exp \left [-\frac{\Phi_0(x,y)}{\varepsilon}-\Phi_1(x,y)\right ]\right ]^{-1}\sim \frac{\sqrt{\det({\bf Z}(x_-,y_-))}}{2\pi \epsilon},
\end{equation}
where ${\bf Z}$ is the Hessian matrix
\begin{equation}
{\bf Z}=\left (\begin{array}{cc} \frac{\partial^2 \Phi_0}{\partial x^2} & \frac{\partial^2 \Phi_0}{\partial x\partial y} \\
 \frac{\partial^2 \Phi_0}{\partial y \partial x} & \frac{\partial^2 \Phi_0}{\partial y^2}\end{array} \right ),
 \end{equation}
and we have chosen $\Phi_0(x_-,y_-)=0=\Phi_1(x_-,y_-)$.

\subsection{Perturbation analysis of the adjoint eigenfunction}
Since $\lambda_0$ is exponentially small, the leading-order equation for the adjoint 1-tensor ${\bm \xi}_0$ is
\begin{equation}
\label{bib}
\epsilon \left [ {\bf V}\frac{\partial}{\partial x}+\widetilde{\bf V}\frac{\partial}{\partial y}\right ]\circ{\bm \xi}_0+{\bf A}^T\circ{\bm \xi}_0=0,
\end{equation}
supplemented by the absorbing boundary conditions (with $({\bm \xi}_0)_{n_x,n_y}\equiv \xi_{n_x,n_y}$)
\begin{equation}
\label{bcsep}
\xi_{n_x,n_y}(x,y)=0,\quad (x,y)\in \Gamma
\end{equation}
for all $(n_x,n_y)$ such that 
\begin{equation}
\label{ls2}
(v(x,n_x,n_y),\widetilde{v}(y,n_x,n_y))\cdot \widehat{\bf s}>0.
\end{equation}
Following along similar lines to \cite{Newby12}, we introduce a new coordinate system $\tau=\tau(x,y),\sigma=\sigma(x,y)$ such that $\tau$ parameterises the separatrix $(x,y)\in \Gamma$ and $\sigma$ is a local coordinate along the normal $\hat{\bf s}$ of $\Gamma$. We scale $\sigma$ so that $(\partial x/\partial \sigma,\partial y/\partial \sigma)=\hat{\bf s}$ at $\sigma=0$. Equation (\ref{bib}) becomes
\begin{equation}
\label{bib2}
\epsilon \left [ {\bf V}_{\tau}\frac{\partial}{\partial \tau}+{\bf V}_{\sigma}\frac{\partial}{\partial \sigma}\right ]\circ {\bm \xi}_0+{\bf A}^T\circ {\bm \xi}_0=0,
\end{equation}
where
\begin{equation}
\label{vsig}
{\bf V}_{\tau}=\frac{\partial \tau}{\partial x}{\bf V}+\frac{\partial \tau}{\partial y}\widetilde{\bf V},\quad {\bf V}_{\sigma}=\frac{\partial \sigma}{\partial x}{\bf V}+\frac{\partial \sigma}{\partial y}\widetilde{\bf V},
\end{equation}
and all terms are rewritten as functions of $\tau,\sigma$. Thus, ${\bf A}(\sigma,\tau)={\bf A}(x(\sigma,\tau),y(\sigma,\tau))$ etc. As a first attempt at obtaining an approximation for ${\xi}_0$, we  introduce a boundary layer around $\Gamma$ by setting $\sigma=\epsilon z$ and ${\bf Q}(z,\tau)={\bm \xi}_0(\epsilon z,\tau)$. To leading-order, equation (\ref{bib2}) becomes
\begin{equation}
\label{Q2}
\left [ {\bf V}_{\sigma}(0,\tau)\frac{\partial}{\partial z}\right ]\circ{\bf Q}(z,\tau)+{\bf A}^T(0,\tau)\circ{\bm Q}(z,\tau)=0.
\end{equation}
The inner solution has to be matched with the outer solution ${\bm \xi}_0={\bf 1}$, which means
\begin{equation}
\lim_{z\rightarrow \infty}{\bf Q}(x,\tau)={\bf 1}.
\end{equation}
We now introduce the eigenfunction expansion (cf. equation (\ref{Qz}))
\begin{equation}
\label{Qz2}
{\bf Q}(z,\tau)=c_0(\tau){\bf 1}+\sum_{j\geq 1}c_j(\tau){\bm \zeta}_j(0,\tau)\e^{\mu_j(0,\tau)z},
\end{equation}
where ${\bf 1}$ has a zero eigenvalue, and
\begin{equation}
\label{zeta2}
\mu_j(\sigma,\tau){\bf V}_{\sigma}(\sigma,\tau)\circ{\bm \zeta}_j(\sigma,\tau)+{\bf A}^T(\sigma,\tau)\circ{\bm \zeta}_j(\sigma,\tau)=0,\quad j \neq 0.
\end{equation}
In order that the solution remain bounded as $z\rightarrow \infty$ and fixed $\tau$, we require that $c_j(\tau)=0$ if $\mu_j(0,\tau)>0$. Suppose that the boundary conditions (\ref{bcsep}) for fixed $\tau$ generate a system of linear equations for the unknown coefficients $c_j(\tau)$ of codimension $k$. One of the coefficients is determined by matching the outer solution, which suggests that there are $k-1$ positive eigenvalues for each $\tau$. The eigenvalues are ordered so that for each $\tau$, $\mu_j(0,\tau)>0$ for $j>k-2$.

Analogous to the one-population model, an additional eigenvalue $\mu_1$, say, vanishes at the saddle point $(0,\tau_*)$ on the separatrix. In order to shows this, suppose that
\begin{equation}
\mu_1 =\frac{{\mathbf 1}^T\circ[{\mathcal P}_x {\bf V}+{\mathcal P}_y\widetilde{\bf V}]\circ {\bm \zeta}_1}{{\mathbf 1}^T\circ{\bf V}_{\sigma}\circ {\bm \zeta}_1}=\frac{\partial \Phi_0}{\partial \sigma},
\end{equation}
where the last expression follows from equations (\ref{Psig}) and (\ref{vsig}).
Substitution into equation (\ref{zeta2}) for $j=1$ then gives
\begin{equation}
[{\bf A}^T+{\mathcal P}_x{\bf V}+{\mathcal P}_y\widetilde{\bf V}]\circ{\bm \zeta}_1=0,
\end{equation}
which has the unique solution ${\bm \zeta}_1={\bm \eta}$, the adjoint of ${\bf R}_0$. Since ${\mathcal P}_x$ and ${\mathcal P}_y$ vanish at $(0,\tau_*)$ and ${\bf V}_{\sigma}\circ {\bm \eta}\neq 0$, it follows that $\mu_1(0,\tau_*)=0$. Hence, the solution at $\tau_*$ has to include a secular term involving the generalized eigenvector ${\bm \zeta}_0$, where
\begin{equation}
{\bf A}^T(0,\tau_*)\circ{\bm \zeta}_0=-{\bf V}_{\sigma}(0,\tau_*)\circ{\bf 1}=-\hat{\bf s}\cdot ({\bf V},\widetilde{\bf V})\circ {\bf 1}.
\end{equation}
The Fredholm alternative theorem ensures that a solution exists, since ${\bm \rho}(0,\tau_*)$ is the left null 1-tensor of ${\bf A}^T(0,\tau_*)$ and ${\bm \rho}(0,\tau_*)\circ{\bf V}(0,\tau_*)={\bm \rho}(0,\tau_*)\circ \widetilde{\bf V}(0,\tau_*)=0$. More explictly, setting $[{\bm \zeta}_0]_{n_x,n_y}=\zeta(n_x,n_y)$, we have
\begin{eqnarray}
&&F(x)\zeta(n_x+1,n_y)+F(y)\zeta(n_x,n_y+1)+n_x\zeta(n_x-1,n_y)+n_y\zeta(n_x,n_y-1)\nonumber \\
&&\quad -[F(x)+F(y)+n_x+n_y]\zeta(n_x,n_y)\nonumber \\
&& \quad =s_x[x-w_{EE}n_x+w_{EI}n_y]+s_y[y-w_{IE}n_x+w_{II}n_y].
\end{eqnarray}
This has a solution of the form $\zeta_{n_x,n_y}={\mathcal A}_xn_x+{\mathcal A}_yn_y$, with the coefficients ${\mathcal A}_x,{\mathcal A}_y$ determined by canceling linear terms in $n_x,n_y$. Thus
\begin{equation}
\zeta_{n_x,n_y}=[w_{EE}s_x+w_{IE}s_y]n_x-[w_{EI}s_x+w_{II}s_y]n_y.
\end{equation}
Given ${\bm \zeta}_0$, equation (\ref{Qz2}) becomes
\begin{equation}
\label{Qz3}
{\bf Q}(z,\tau_*)=c_0(\tau_*){\bf 1}+c_1(\tau_*)({\bm \zeta}_0-z{\bf 1})+\sum_{j\geq 2}c_j(\tau_*){\bm \zeta}_j(0,\tau_*)\e^{\mu_j(0,\tau_*)z},
\end{equation}
The presence of the secular term implies that the solution is unbounded so it has to be eliminated using a modified stretch variable $\sigma =\sqrt{\epsilon }z$ \cite{Newby12a,Newby12}. As in the one-population case, we find that
\begin{equation}
\label{c12}
c_1(\tau_*)\sim -\sqrt{\frac{2\epsilon |\partial_{\sigma}\mu_1(0,\tau_*)|}{\pi}}.
\end{equation}

\subsection{Principal eigenvalue}

We now return to the expression for the principal eigenvalue $\lambda_0$ given by equation (\ref{lam02}). Simplifying the denominator by using the outer solution ${\bm \xi}_0\sim {\bf 1}$  and using the WKB approximation of $\vphi$, equation (\ref{quiz}), gives
\begin{equation}
\lambda_0={\mathcal N}\int_{\Gamma}\e^{-\Phi_1(x,y)}\e^{-\Phi_0(x,y)/\epsilon} \left ({\bm \xi}_0^T\circ{\bf V}\circ {\bf R}_0 ,\, {\bm \xi}_0^T\circ \widetilde{\bf V}\circ {\bf R}_0 \right )\cdot \widehat{\bf n}ds .
\end{equation}
Changing to the $(\sigma,\tau)$ coordinate system and evaluating the line integral by applying Laplace's method around the saddle point $(0,\tau^*)$ then gives
\begin{eqnarray}
\label{eq:2}
\lambda_0&\sim& {\mathcal N}B(\tau_*)c_1(\tau_*)\e^{-\Phi_1(0,\tau_*)}\e^{-\Phi_0(0,\tau_*)/\epsilon}\int_{\Gamma} \exp \left (-\frac{1}{2\epsilon} \partial_{\tau\tau}\Phi_0(0,\tau_*)(\tau-\tau_*)^2\right )d\tau,\nonumber \\
&\sim& B(\tau_*)c_1(\tau_*)\e^{-\Phi_1(0,\tau_*)}\e^{-\Phi_0(0,\tau_*)/\epsilon}\sqrt{\frac{2\pi}{\partial_{\tau\tau}\Phi_0(0,\tau_*)}}\frac{\sqrt{\det({\bf Z}(x_-,y_-))}}{2\pi \epsilon}\nonumber \\
&\sim &\frac{1}{\pi} B(\tau_*)\e^{-\Phi_1(0,\tau_*)}\e^{-\Phi_0(0,\tau_*)/\epsilon}\sqrt{\frac{\partial_{\sigma\sigma}\Phi_0(0,\tau_*)\det({\bf Z}(x_-,y_-))}{\partial_{\tau\tau}\Phi_0(0,\tau_*)}},
\end{eqnarray}
where we have used equations (\ref{norm2}), (\ref{Qz3}), (\ref{c12}), and
\begin{equation}
B(\tau_*)=\left ({\bm \xi}_0^T\circ{\bf V}\circ {\bf R}_0(0,\tau_*) ,\, {\bm \xi}_0^T\circ \widetilde{\bf V}\circ {\bf R}_0(0,\tau_*) \right )\cdot \widehat{\bf n}.
\end{equation}

\subsection{Results}
The rays $(x(t), y(t))$ (i.e., solutions to the Hamilton's equations \eqref{eq:4} in the $(x,y)$ plane) have an important physical meaning.  The trajectory of the ray is the most likely trajectory or path leading away from a stable fixed point \cite{Dykman94}.  Under this interpretation, one can describe the stochastic dynamics using the metastable dynamical trajectories (rays) along with deterministic trajectories.
\begin{figure}[htbp]
  \centering
  \includegraphics[width=12cm]{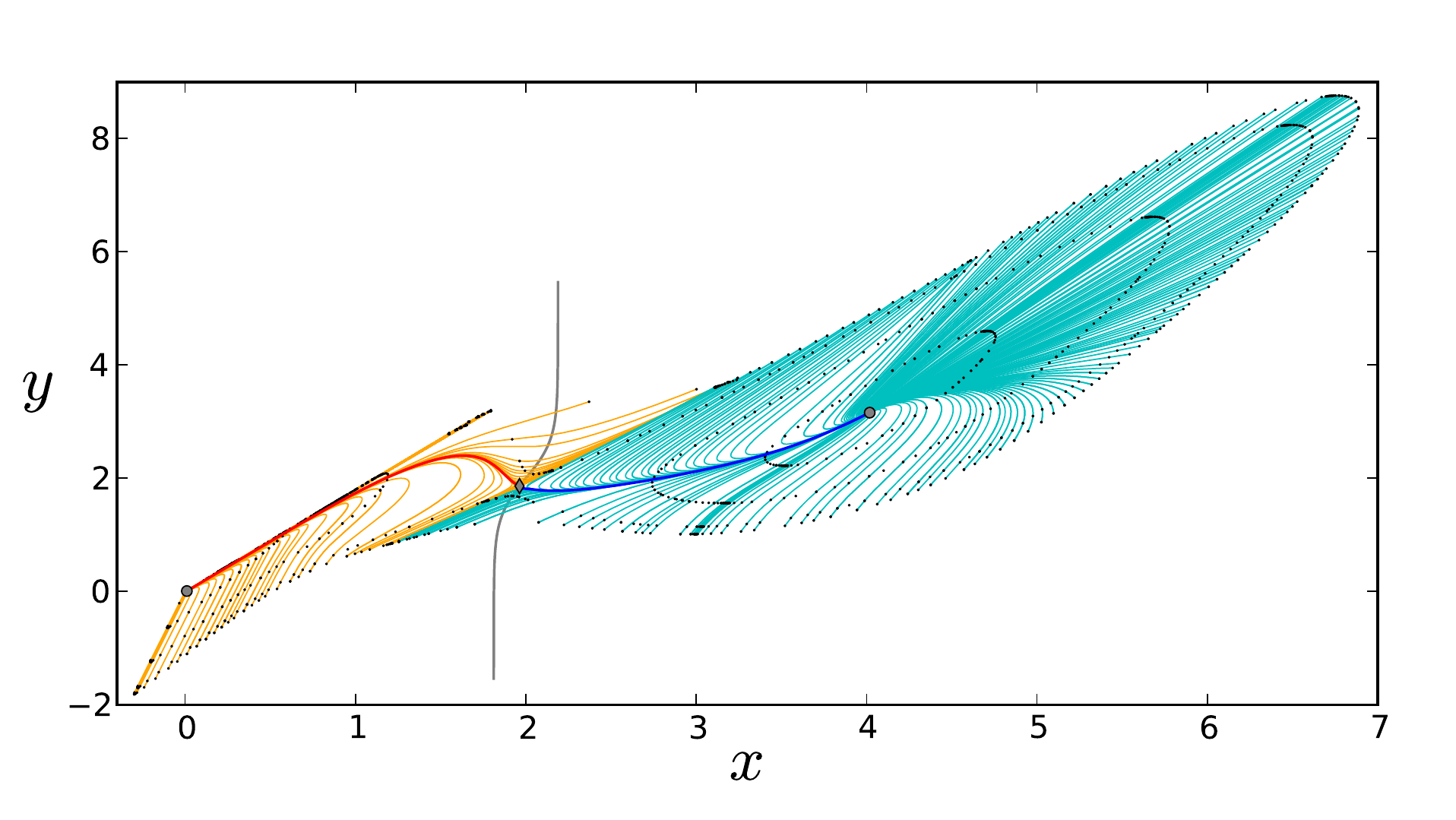}
  \includegraphics[width=12cm]{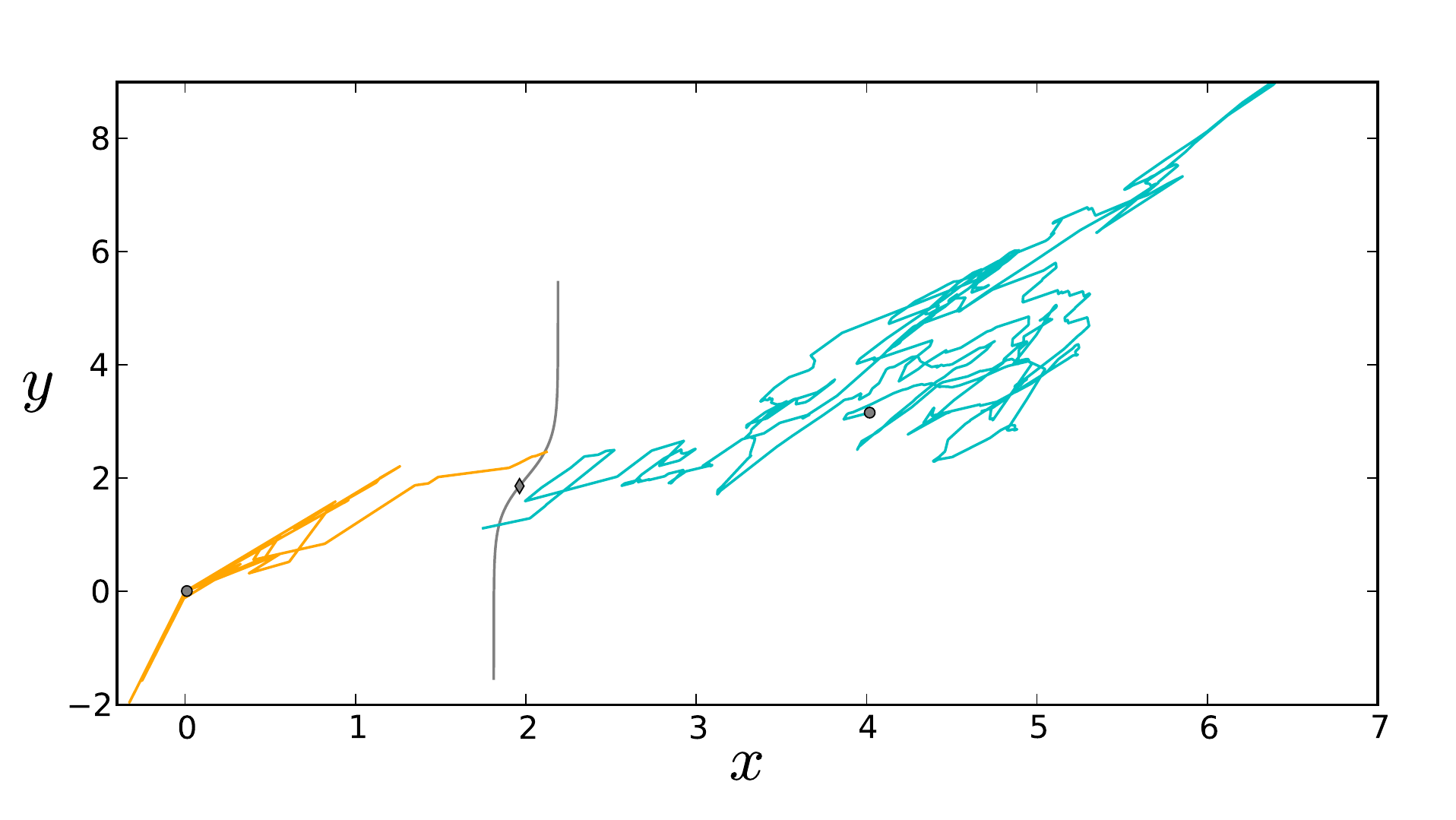}
  \caption{(a) Characteristic paths of maximum likelihood for the 2D model.  Rays originating from the left (right) stable fixed point are shown in orange (cyan), with the ray connecting to the saddle shown in red (blue).  The grey curve is the separatrix $\Gamma$.  Level curves of constant $\Phi(x, y)$ are shown as black dots.  Each ray has four dots for different values of $\Phi(x, y)$.  Rays originating from the left fixed point have dots at $\Phi = 0.1, 0.2, \Phi_{*} + 0.01,   \Phi_{*}+0.02$, and rays originating from the right fixed point have dots at $\Phi = 0.19, 0.23, 0.28, 0.30$,  where $\Phi_{*} = \Phi(x_{*}, y_{*}) = 0.28$.   All rays terminate at $\Phi = \Phi_{*}+0.02$.  (b) Sample trajectories of the two-population velocity jump Markov process, whose associated probability density evolves according to (\ref{CK2}), are computed using the Gillespie algorithm with $\epsilon = 0.05$ and $N\Delta t = 1$. (The maximum likelihood paths are independent of epsilon).  Other parameter values are the same as in Fig.~\ref{fig:bistable2}.}
  \label{fig:rays}
\end{figure}
The rays $(x(t), y(t))$ shown in Fig.~\ref{fig:rays} are obtained by integrating the characteristic equations \eqref{HHx} and \eqref{HHy}.  These trajectories are only valid in one direction: away from the stable fixed points.  The most likely trajectory leading toward stable fixed points are given by deterministic dynamics (see Fig.~\ref{fig:bistable2}).  For parameter values considered in Fig.~\ref{fig:rays}, rays originating from each stable fixed point cover separate regions, so that most likely paths between points in each region are connected by deterministic trajectories starting at the boundary between the two regions.  Note that this boundary is not the separatrix (grey curve).  For example, a trajectory initially at the left fixed point which crosses the separatrix at the saddle would most likely follow a ray toward the saddle and then follow a deterministic trajectory to the right fixed point.  If a trajectory crosses the separatrix away from the saddle, it is most likely to cross the separatrix above the saddle when starting from the left fixed point and below the saddle when starting from the right fixed point (see Fig.~\ref{fig:exit_rays}).  At first glance, this suggests that if the trajectory starts at the left fixed point, say, it is more likely to cross above the saddle, continue along a deterministic trajectory to the right fixed point, and then cross the separatrix below the saddle than it is to directly cross below the saddle.  This is counter to intuition because it would seem more likely for a single rare, metastable crossing event to lead to a point near the separatrix than two rare events occurring in sequence.  However, as shown in \cite{Stein97}, rays can also originate from the saddle point that cross the separatrix in the direction oposite those originating at the stable fixed points.
\begin{figure}[htbp]
  \centering
  \includegraphics[width=10cm]{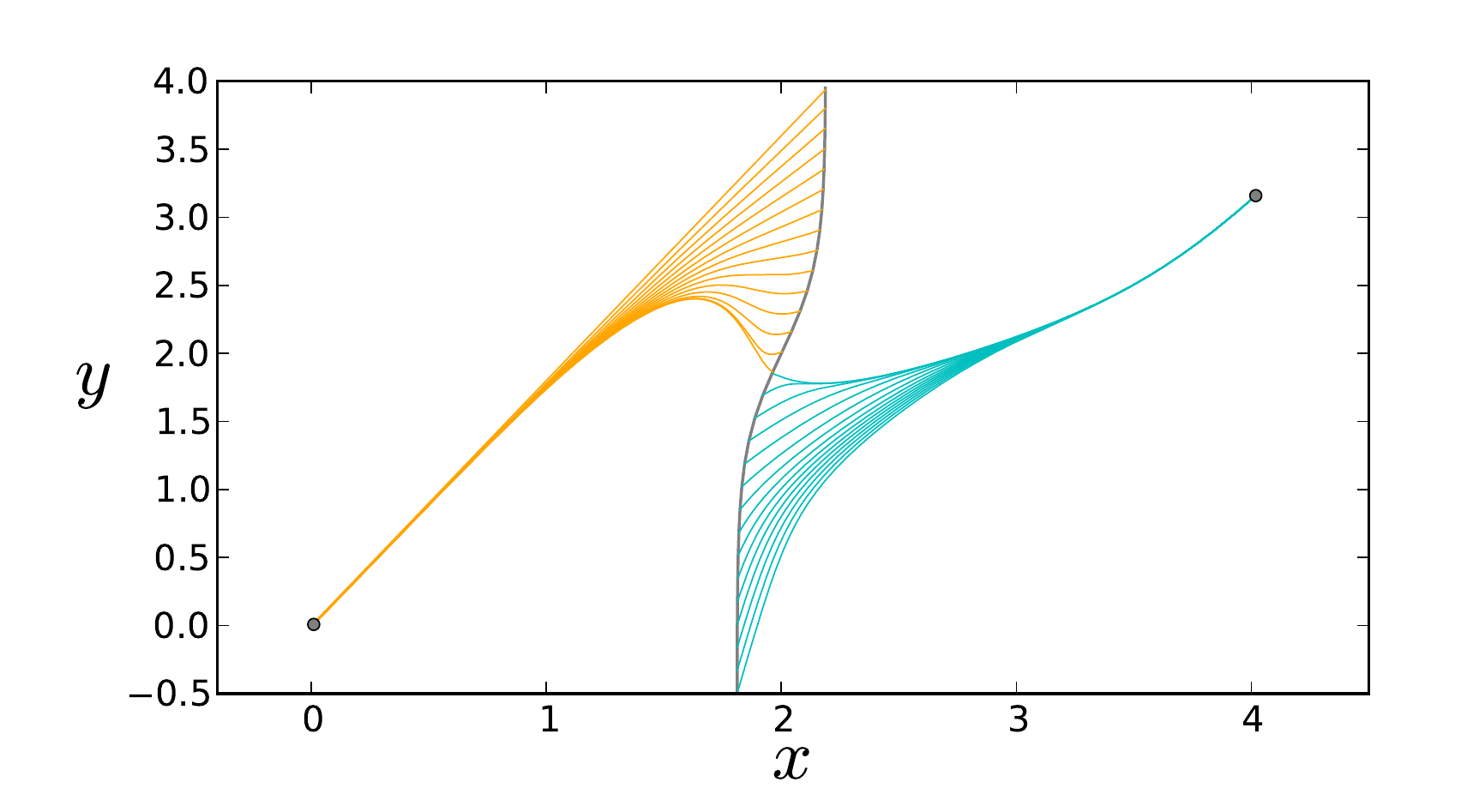}  
  \caption{Maximum-liklihood trajectories crossing the separatrix.}
  \label{fig:exit_rays}
\end{figure}
\begin{figure}[htbp]
  \centering
  \includegraphics[width=10cm]{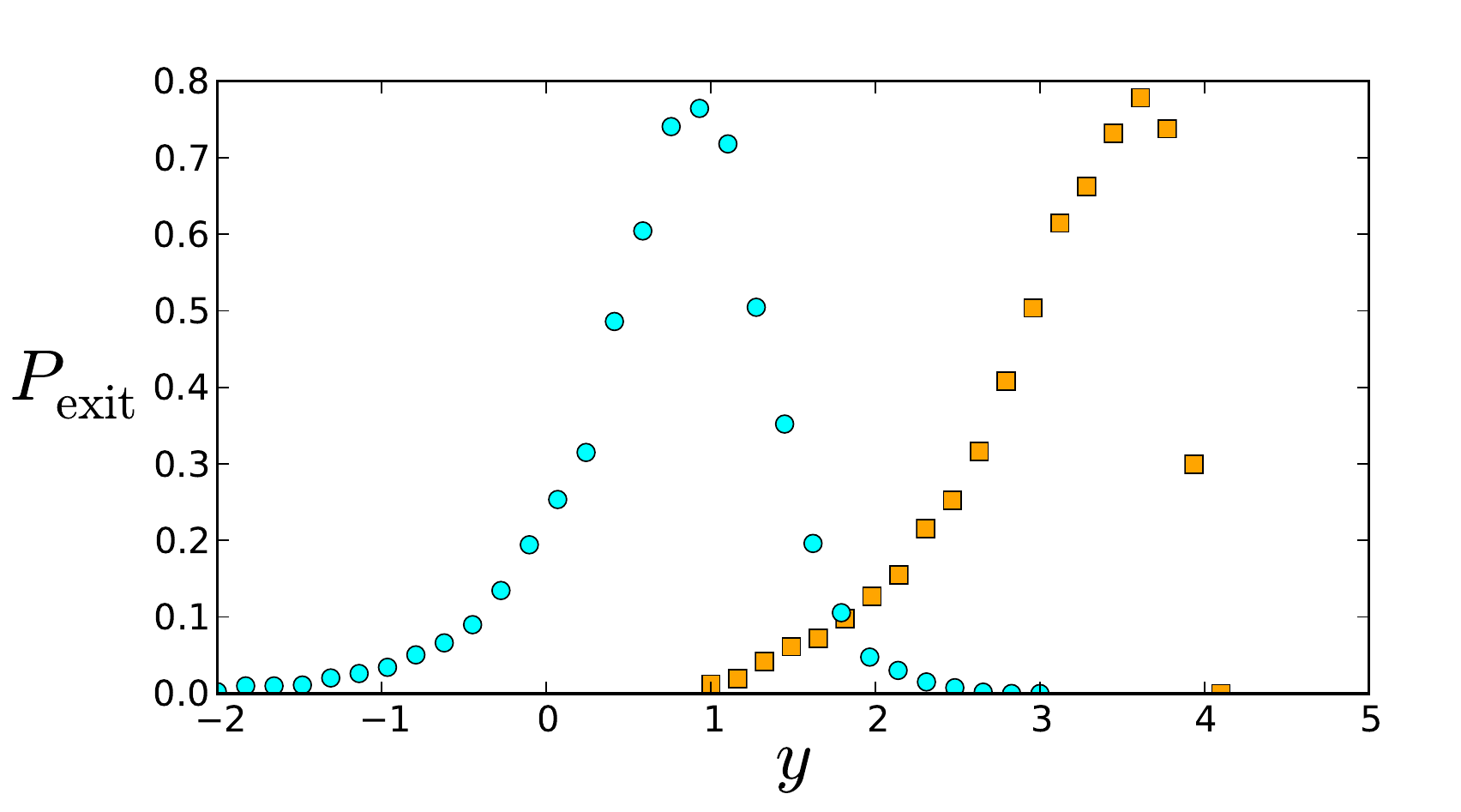}
  \caption{The probability density for the exit point ($y$ coordinate) where the separatrix is crossed by an exiting trajectory.  Results are obtained by $10^{2}$ Monte-Carlo simulation with the same parameters as used in Fig.~\ref{fig:bistable2}, with $\epsilon = 0.08$.  The square symbols show trajectories from the left well, and 'o' symbols show trajectories from the right well.}
  \label{fig:exit_dist}
\end{figure}
In Fig.~\ref{fig:exit_dist}, the probability density function for the $y$ coordinate of the point on the separatrix reached by an exit trajectory is shown for each well (square symbols show the histogram for exit from the left well and likewise, 'o' symbols for the right well).  Each density function is peaked away from the saddle point, showing a phenomena known as saddle point avoidance \cite{Stein97,Schuss10}.    As $\epsilon\to 0$, the two peaks merge at the saddle point.  Although we expect the saddle point to be the most likely exit point since it the point on the separatrix where the potential $\Phi$ takes its minimum value, our results show that this is not necessarily true.

\begin{figure}[htbp]
  \centering
  \includegraphics[width=12cm]{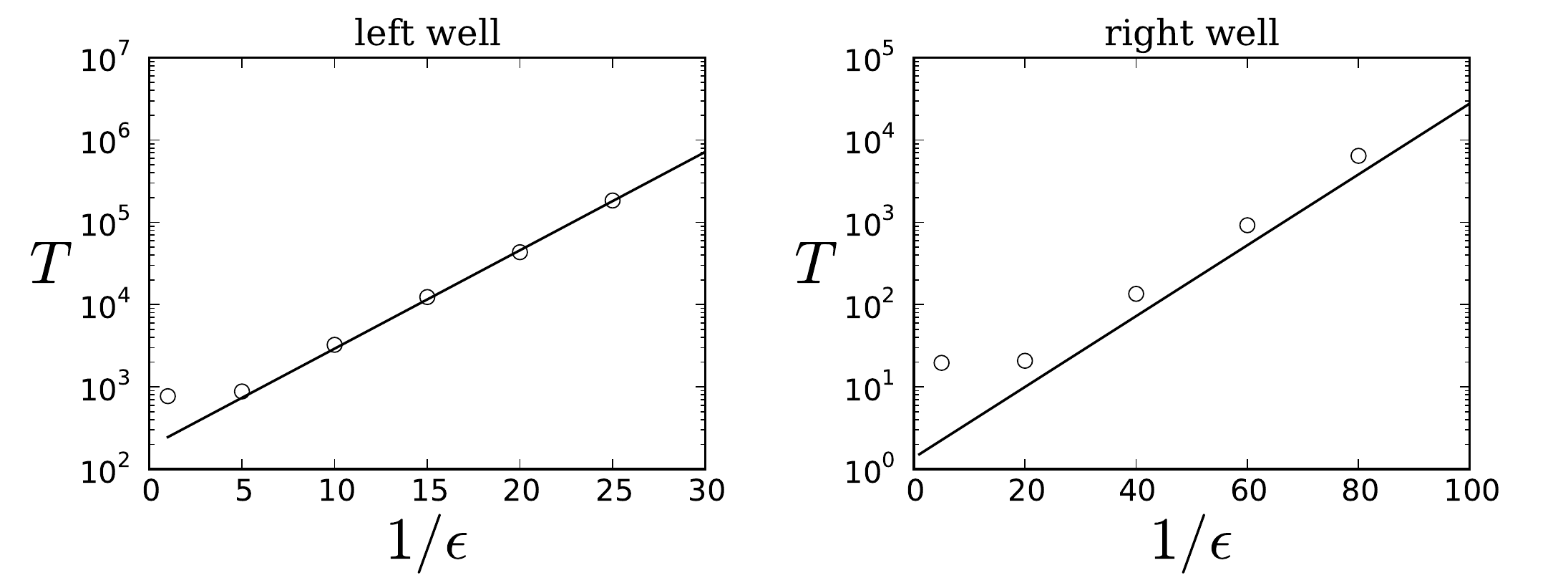}
  \caption{Mean exit time from the left and right well. Parameter values are the same as in Fig.~\ref{fig:bistable2}.  Solid lines show the analytical approximation $T\sim 1/\lambda_{0}$, where $\lambda_{0}$ is given by \eqref{eq:2}, and symbol show $80$ averaged Monte-Carlo simulations.}
  \label{fig:2Dmfpt}
\end{figure}

Even though the most likely exit point is shifted from the saddle, the value of potential at the saddle point still dominates the mean first exit time. In Fig.~\ref{fig:2Dmfpt}, the mean exit time from each of the 2D potential wells (see Fig.~\ref{fig:rays}) is shown.  Solid lines show the analytical approximation $T\sim 1/\lambda_{0}$, where $\lambda_{0}$ is given by \eqref{eq:2}, and symbol show averaged Monte-Carlo simulations.  As in Fig.~\ref{fig:1Dmfpt}, the slope $T$ on a log scale as a function of $1/\epsilon$ is determined by $\Phi$ evaluated at the saddle point.

\section{Discussion}

In this paper we developed a generalization of the neural master equation \cite{Buice07,Bressloff09,Buice09}, based on a velocity jump Markov process that couples synaptic and spiking dynamics at the population level. There were two distinct time-scales in the model, corresponding to the relaxation times $\tau$ and $\tau_a$ of the synaptic and spiking dynamics, respectively. In the limit $\tau\rightarrow 0$, we recovered the neural master equation operating in a Poisson-like regime, whereas in the limit $\tau_a\rightarrow 0$ we obtained determistic mean field equations for the synaptic currents. Hence, one additional feature of our model is that it provides a prescription for constructing a stochastic population model that reduces to a current-based model, rather-than an activity-based model, in the mean-field limit. 

We focused on the particular problem of escape from a metastable state, for which standard diffusion--like approximations break down. We showed how WKB methods and singular perurbation theory could be adapted to solve the escape problem for a velocity jump Markov process, extending recent studies of stochastic ion channels. For concreteness, we assumed that the network operated in the regime $\tau_a/\tau=\epsilon \ll1$, which meant that transitions between different discrete states of population spiking activity were relatively fast. In this regime, the thermodynamic limit $N\rightarrow \infty$ was not a mean-field limit, rather it simplified the analysis since the quasi-steady-state density was Poisson. It would be interesting to consider other parameter regimes in subsequent work. First, we could model the discrete Markov process describing the spiking dynamics using the Bressloff version of the master equation \cite{Bressloff09}. There would then be two small parameters in the model, namely $\epsilon$ and $1/N$, so one would need to investigate the interplay between the system size expansion for large but finite $N$ and the quasi-steady--state approximation for small $\epsilon$. Another possible scenario (though less plausible physiologically speaking) would be fast synaptic dynamics with $\tau \ll \tau_a$. In this case, mean-field equations are obtained in the thermodynamic limit. Finally, it would be interesting to extend our methods to analyze the effects of noise when the underlying deterministic sytstem exhibite more complicated dynamics such as limit cycle oscillations. As we commented in the main text, the two-population model of excitatory and inhibitory neurons is a canonical circuit for generating population-level oscillations.

Finally, it is important to emphasize that the neural master equation and its generalizations are phenomenological models of stochastic neuronal population dynamics. Although one can give a heuristic derivation of these models \cite{Bressloff12}, there is currently no sytematic procedure for deriving them from physiologically-based microscopic models, except in a few special cases. Nevertheless, stochastic hybrid models are emerging in various applications within neuroscience, so that the analytical techniques presented in this paper are likely to be of increasing importance.
\bibliography{nft}
\bibliographystyle{siam}

\end{document}